\documentclass[10pt,twocolumn,showpacs,showkeys,preprintnumbers,amssymb,aps,superscriptaddress,pre]{revtex4-2}
\usepackage{color}
\usepackage{amsmath}
\usepackage{graphicx}
\usepackage{bm}
\usepackage{wrapfig}
\usepackage{subfigure}
\usepackage{epsfig}
\usepackage{epstopdf}
\usepackage{relsize}

\begin{document}

\preprint{APS/123-QED}
\title{Unconventional spin frustration due to two competing ferromagnetic interactions \\ of a spin-1/2 Ising-Heisenberg model on martini and martini-diced lattice}

\author{Hamid Arian Zad}
\affiliation{A.I. Alikhanyan National Science Laboratory, 0036, Yerevan, Armenia}
\affiliation{ICTP, Strada Costiera 11, I-34151 Trieste, Italy}
\affiliation{Department of Theoretical Physics and Astrophysics, Faculty of Science, P. J. S\v{a}f{\' a}rik University, Park Angelinum 9, 041 54 Ko\v{s}ice, Slovak Republic}
\author{Jozef Stre\v{c}ka}
\email{Corresponding author: jozef.strecka@upjs.sk}
\affiliation{Department of Theoretical Physics and Astrophysics, Faculty of Science, P. J. S\v{a}f{\' a}rik University, Park Angelinum 9, 041 54 Ko\v{s}ice, Slovak Republic}

\date{\today}

\begin{abstract}
The spin-1/2 Ising-Heisenberg model on martini and martini-diced lattice is exactly solved using a star-triangle transformation, which affords an exact mapping correspondence to an effective spin-1/2 Ising model on a triangular lattice. The ground-state phase diagram of both investigated quantum spin models display two spontaneously ordered ferromagnetic phases and one macroscopically degenerate disordered phase. In contrast to a classical ferromagnetic phase where the spontaneous magnetization of the Ising as well as Heisenberg spins acquire fully saturated values the spontaneous magnetization of the Heisenberg spins is subject to a quantum reduction to one-third of its saturated value within a quantum ferromagnetic phase. The spontaneous magnetization and logarithmic divergence of the specific heat as the most essential features of both ferromagnetic phases disappear whenever the investigated quantum spin model is driven to the highly degenerate disordered phase. The disordered phase with nonzero residual entropy originates either from a geometric spin frustration caused by antiferromagnetic interactions or more strikingly it may also alternatively arise from a competition of the ferromagnetic Ising and Heisenberg interactions of easy-axis and easy-plane type, respectively. All three available ground states coexist together at a single triple point, around which anomalous magnetic and thermodynamic behavior can be detected.
\end{abstract}

\maketitle

\section{Introduction}

Two-dimensional (2D) quantum Heisenberg spin models traditionally attract a great deal of attention as they may eventually display at zero temperature a peculiar quantum spin liquid with unconventional topological order instead of a conventional spontaneous ordering \cite{ric04,mis04,bal10}. A precise nature of the exotic quantum phases is often subject of controversial debate, because unconventional quantum spin orders are extremely fragile even with respect to small perturbations, which is why they can be captured just by rigorous analytical or numerical methods \cite{ric04,mis04,bal10}. An unbiased treatment of 2D quantum Heisenberg spin models is however connected with formidable mathematical difficulties that are hard to overcome. One possible way how to avoid this tremendous obstacle is to replace some quantum Heisenberg spins through their classical Ising counterparts within the so-called Ising-Heisenberg spin models, which are exactly tractable by combining generalized mapping transformations \cite{fis59,roj09,str10} with the transfer-matrix method \cite{kra41,bax82}.

In spite of their considerably simplified nature, exactly solvable Ising-Heisenberg spin models represent appealing issue to deal with as they provide beneficial playground for investigating cooperative quantum phenomena without any unbiased approximation \cite{str02,str08,yao08,cis13,zho18,roj19,gal20,gal21}. It is worthwhile to remark that 2D Ising-Heisenberg models additionally allow existence of a spontaneous long-range order not only at absolute zero temperature, because the Mermin-Wagner theorem \cite{mer66} does not apply for these 2D lattice-statistical models due to an incorporation of the Ising spins. The first exactly solved 2D spin-1/2 Ising-Heisenberg model on a doubly decorated square lattice displays for instance either a classical ferromagnetic or a quantum antiferromagnetic long-range order at sufficiently low temperatures, whereby the quantum antiferromagnetic ordering peculiarly comes from two competing ferromagnetic interactions \cite{str02}. The competition between the Ising and Heisenberg interactions might be also at origin of a more complex critical behavior including reentrant phase transitions \cite{str06,can10,gal11,gal19} or even weak-universal criticality \cite{str09}. It is worth mentioning, moreover, that a spontaneous long-range order may absent in 2D Ising-Heisenberg models being subject to a geometric spin frustration though the relevant residual entropy is generally lower in comparison with the Ising counterparts due to local quantum fluctuations \cite{str08,yao08,cis13}.   

Although the classical-quantum spin models incorporating exchange-coupled Ising and Heisenberg spins might seem at first glance to be of purely theoretical interest without connection to any real-world system, several paradigmatic examples of the Ising-Heisenberg spin models have proved their significance when capturing underlying magnetic properties of a few selected magnetic compounds such as Cu(3-Clpy)$_2$(N$_3$)$_2$ \cite{str05}, [(CuL)$_2$Dy][Mo(CN)$_8$] \cite{heu10,bel14}, [Fe(H$_2$O)(L)][Nb(CN)$_8$][Fe(L)] \cite{sah12}, Dy(NO$_3$)(DMSO)$_2$Cu(opba)(DMSO)$_2$ \cite{str12,han13,tor18}, \{Dy(hfac)$_2$(CH$_3$OH)\}$_2$\{Cu(dmg)(Hdmg)\}$_2$ \cite{str20} and [CuMn(L)][Fe(bpb)(CN)$_{2}$] $\cdot$ ClO$_{4}$ $\cdot$ H$_{2}$O \cite{sou20}. The search for exactly solvable Ising-Heisenberg models is thus not only of pure theoretical interest, but it could be also relevant for a targeted design of appropriate magnetic compounds composed of highly anisotropic (e.g. Dy$^{3+}$) and almost isotropic (e.g. Cu$^{2+}$) magnetic ions affording experimental representatives of the Ising and Heisenberg spins, respectively.  

In the present article we will exactly solve the spin-1/2 Ising-Heisenberg model on the martini and martini-diced lattice by employing a generalized version of the star-triangle transformation \cite{ons44}, which proves a rigorous mapping correspondence to an effective spin-1/2 Ising model on a triangular lattice \cite{wan50,hou50,new50}. It is noteworthy that the martini lattice can be derived from a hexagonal lattice by replacing each second site of this bipartite lattice with a triangle, while the martini-diced lattice can be descended from a diced lattice by replacing each its three-coordinate site with a triangle (see Fig. \ref{fig:TIT3}). The previous studies have reported only site and bond percolation thresholds of the martini lattice \cite{scu06,zif06}, while the phase diagram of the spin-1/2 Ising-Heisenberg model on a martini-diced lattice was announced in our preliminary brief report \cite{kis10}. In the following, we will explore in detail  the ground-state and finite-temperature phase diagrams, spontaneous magnetization and specific heat of the spin-1/2 Ising-Heisenberg model on martini and martini-diced lattices. 

This paper is organized as follows. In Sec. \ref{model} we will thoroughly describe both investigated lattice-statistical spin models and basic steps of their exact analytical solutions. In Sec. \ref{results} we will discuss the most interesting results for the ground-state and finite-temperature phase diagrams, spontaneous magnetization, entropy and specific heat order parameters and thermodynamics of the models.
Finally, the most important results of our study are summarized in Sec. \ref{conclusions}. 
 
\section{Model and its exact solution}
\label{model}

\begin{figure*}
\begin{center}
  \centering
\includegraphics[scale=0.52,trim=50 20 00 50, clip]{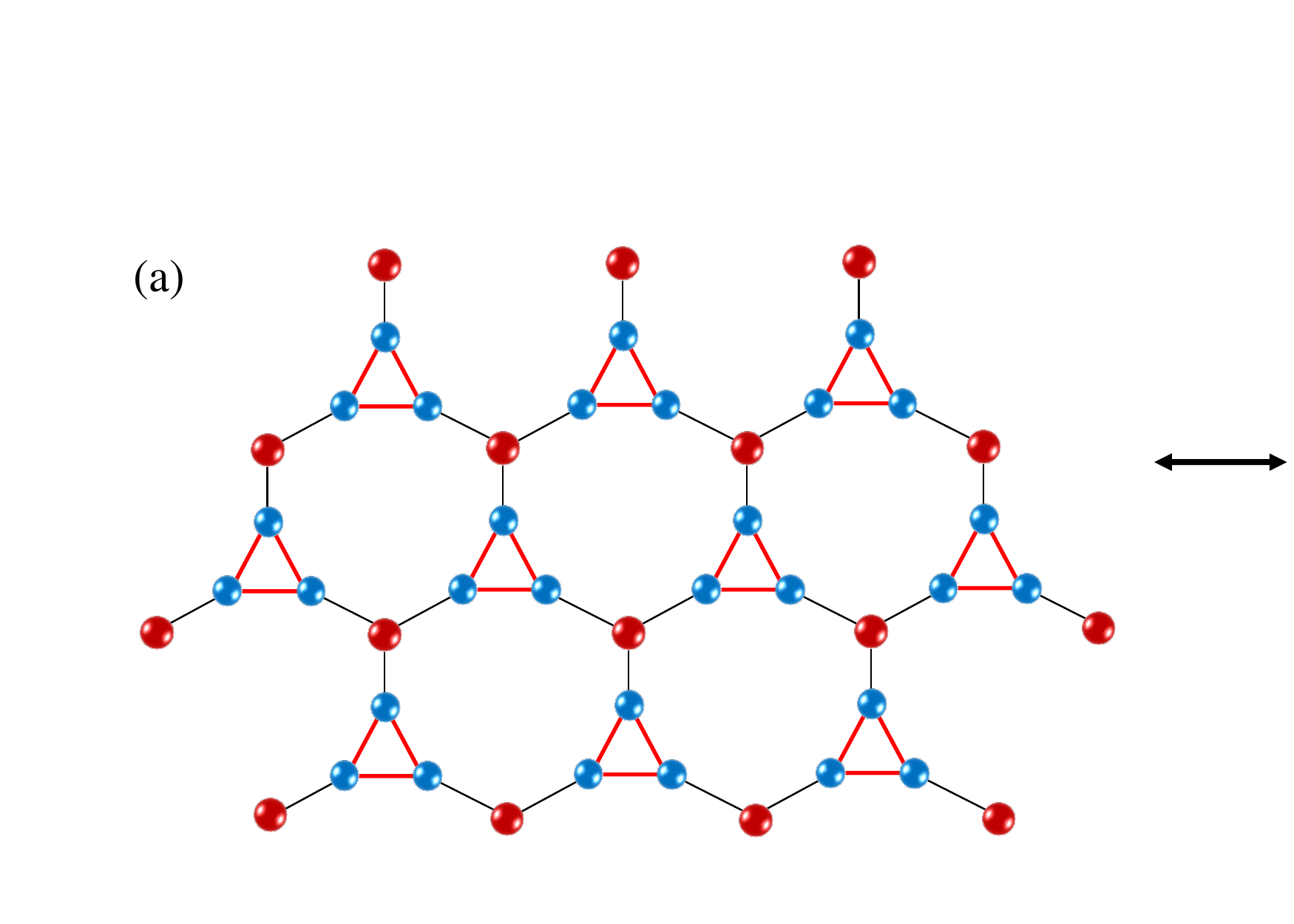}
\includegraphics[scale=0.52,trim=20 50 50 50, clip]{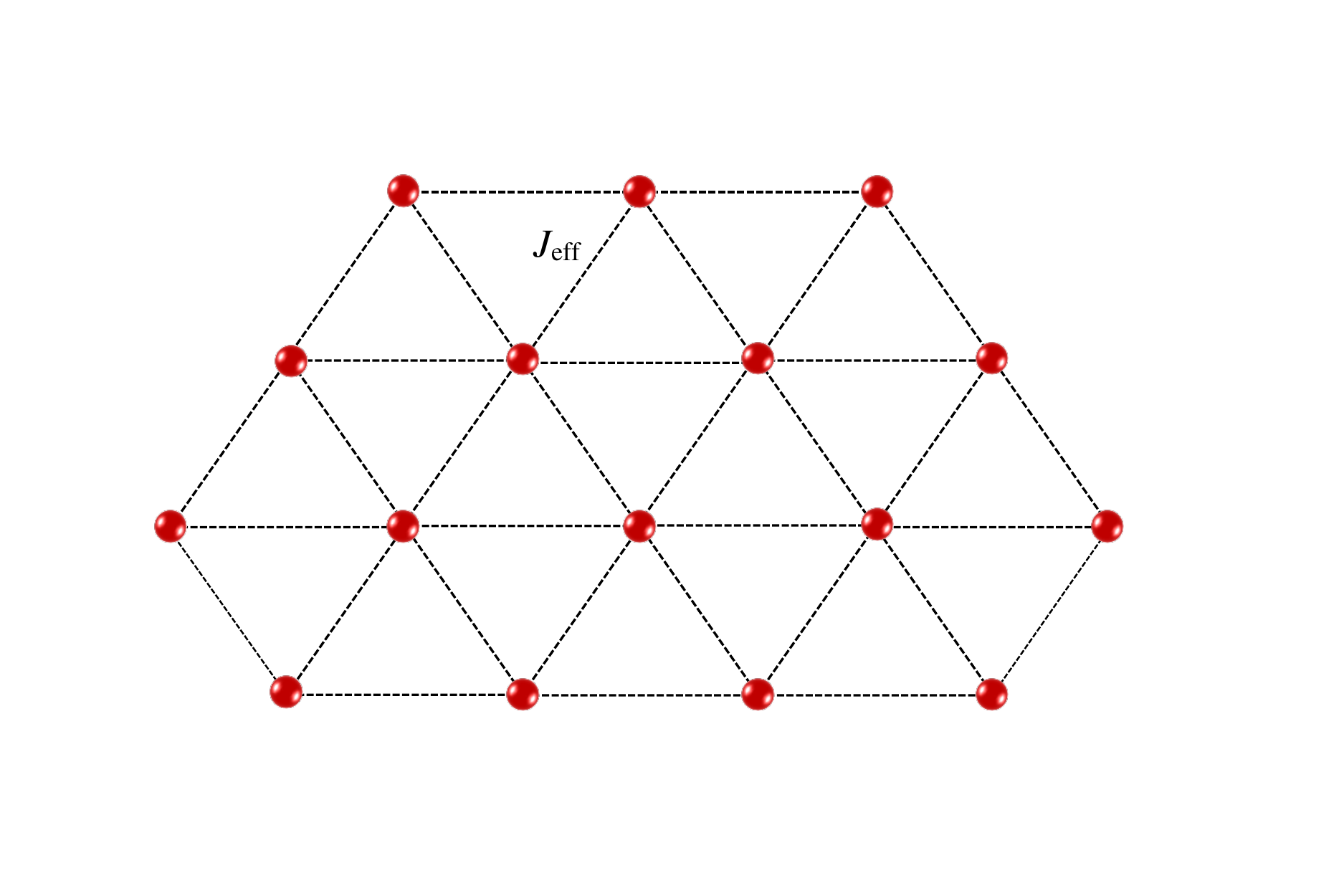}
\\
\includegraphics[scale=0.52,trim=50 20 00 70, clip]{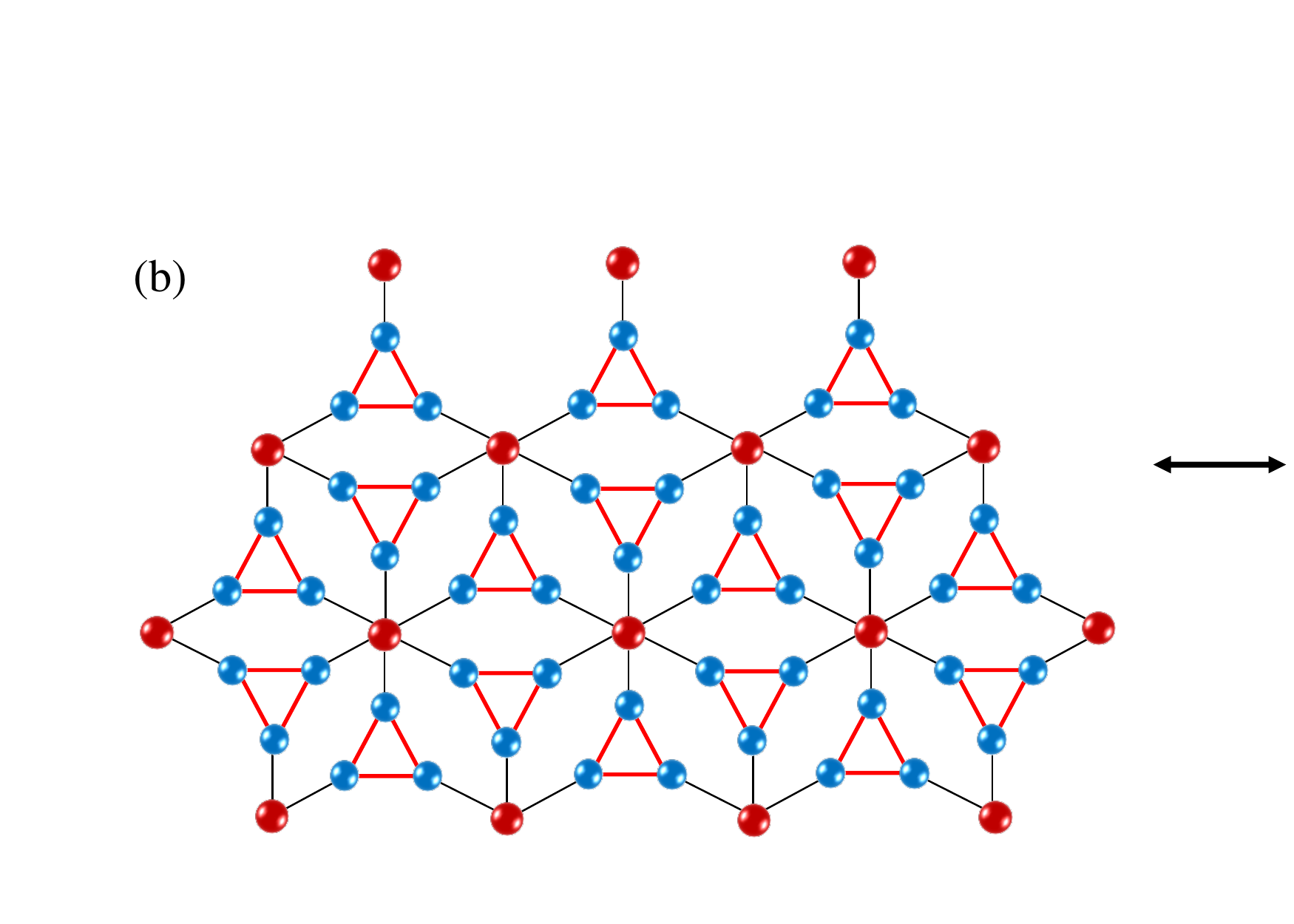}
\includegraphics[scale=0.52,trim=20 50 50 70, clip]{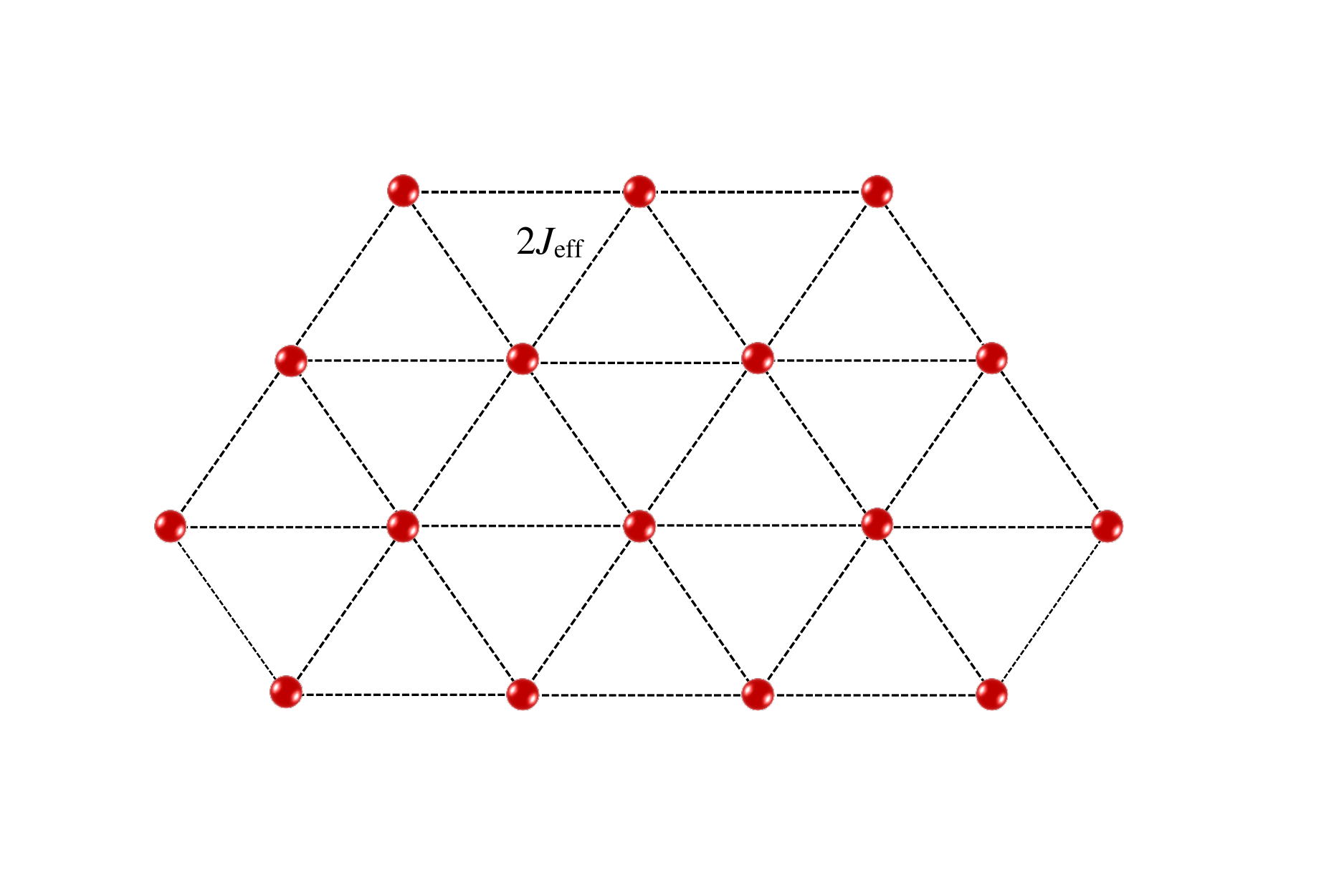}
\\
\includegraphics[scale=0.55,trim=30 80 40 80, clip]{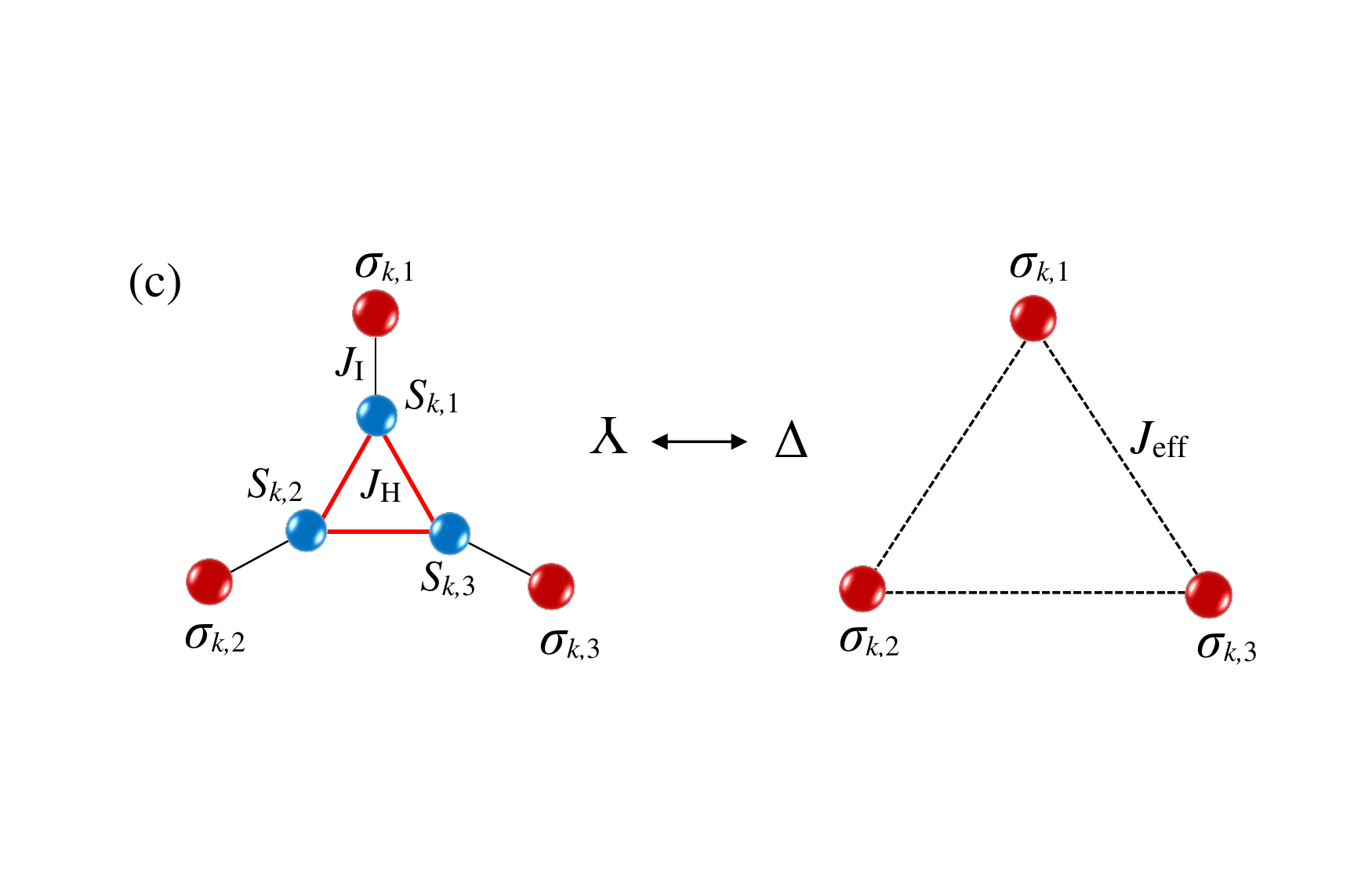}
\includegraphics[scale=0.55,trim=20 80 40 80, clip]{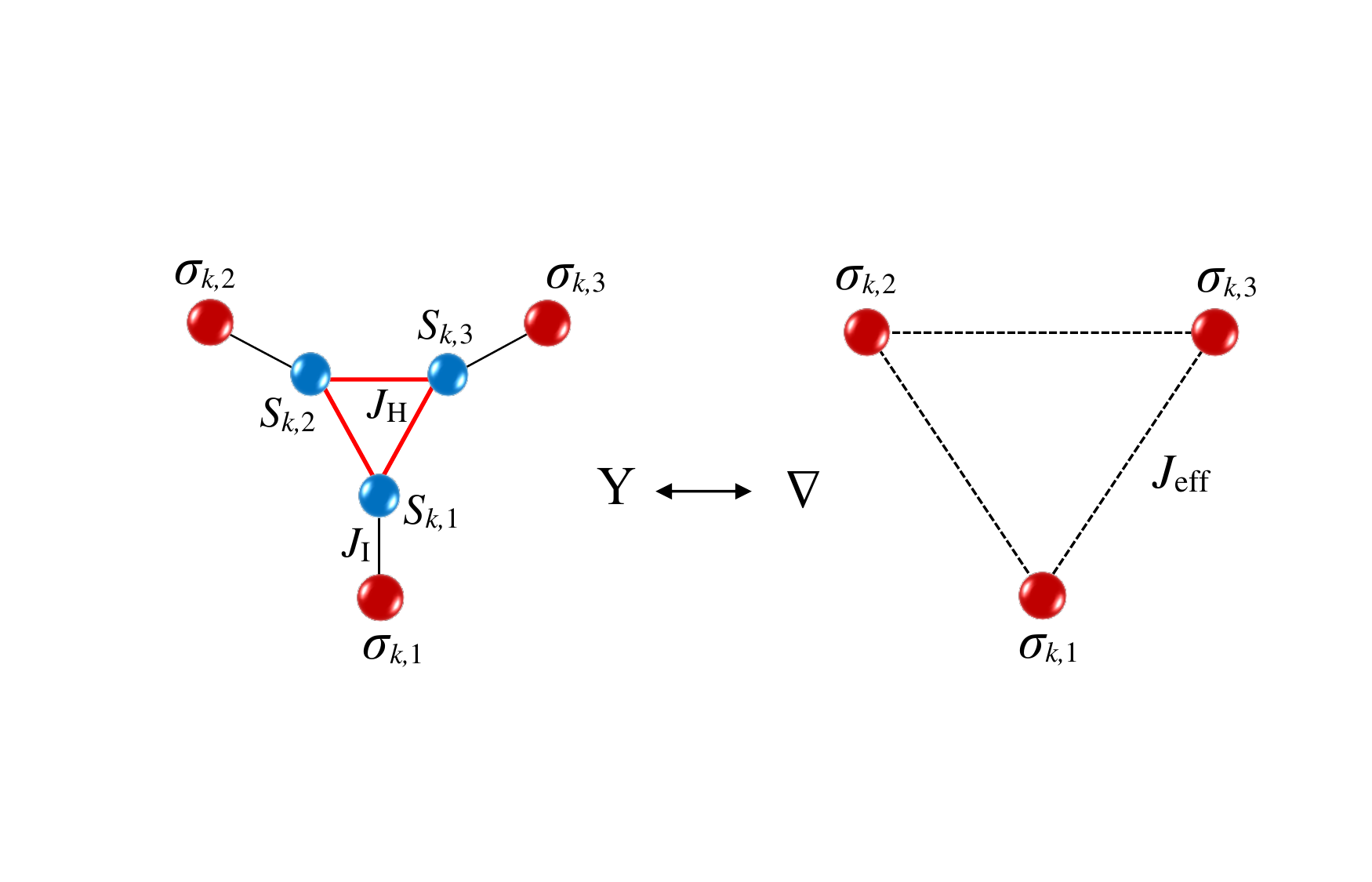}
\caption{A schematic illustration of the spin-1/2 Ising-Heisenberg model on the martini lattice with $\gamma=1$ [Fig. \ref{fig:TIT3}(a) left] and the martini-diced lattice with $\gamma=2$ [Fig. \ref{fig:TIT3}(b) left] together with the corresponding spin-1/2 Ising model on a  triangular lattice [Fig. \ref{fig:TIT3}(a)-(b) right] with the effective interaction $J_\text{eff}$ and $2J_\text{eff}$, respectively. Blue circles denote lattice positions of the Heisenberg spins $S_k$ coupled together via XXZ interaction $J_\text{H}$ (red solid lines), while red circles demarcate lattice sites of the Ising spins $\sigma_k$ coupled to adjacent Heisenberg spins via the Ising interaction $J_\text{I}$. (c) Local star-triangle mapping transformations used for the martini unit cell.}
\label{fig:TIT3}      
\end{center}
\end{figure*}

Let us proceed to a definition of the spin-1/2 Ising-Heisenberg model on the martini and martini-diced lattice given by the total Hamiltonian
\begin{equation}
\label{HamiltonianTIT}
\begin{array}{lcl}
\hat{\cal{H}} = -J_\text{H} \!\sum\limits_{\langle i,j\rangle} \!\! \big[\Delta\big(\hat{S}_i^x\hat{S}_j^x+\hat{S}_i^y\hat{S}_j^y\big)+\hat{S}_i^z\hat{S}_j^z\big]- J_\text{I} \! \sum\limits_{\langle i,k\rangle} \!\! \hat{S}_i^z\hat{\sigma}_k^z,  
\end{array}
\end{equation}
where $\hat{S}_i^{\alpha}$  ($\alpha= x,y,z$) and $\hat{\sigma}_i^z$ are spatial components of the spin-1/2 operator assigned to the Heisenberg and Ising spins schematically depicted in Fig. \ref{fig:TIT3} by blue and red circles, respectively. The coupling constant $J_\text{H}$ stands for the XXZ interaction between the nearest-neighbor Heisenberg spins, $\Delta$ is a spatial anisotropy in this interaction, while the coupling constant $J_\text{I}$ denotes the Ising interaction between adjacent Ising and Heisenberg spins.

The general form of the Hamiltonian (\ref{HamiltonianTIT}) for the spin-1/2 Ising-Heisenberg model on the martini and martini-diced lattice can be alternatively rewritten as
 \begin{equation}
\begin{array}{lcl}
  \hat{\cal H} =\sum\limits_{k=1}^{\gamma N} \hat{\mathcal{H}}_k,
\end{array}
\end{equation}
where $N$ denotes the total number of the Ising spins, $\gamma N$ labels the total number of martini-type six-spin clusters shown in Fig. \ref{fig:TIT3}(c), $\gamma = 1$ for the traditional martini lattice shown in Fig. \ref{fig:TIT3} (a) and $\gamma = 2$ for the related martini-diced lattice displayed in Fig. \ref{fig:TIT3} (b). The cluster Hamiltonian $\hat{\mathcal{H}}_k$ of a single martini-type six-spin cluster [see Fig. \ref{fig:TIT3}(c)] can be defined as
\begin{equation}\label{oneblockH}
\begin{array}{ll}
  \hat{\mathcal{H}}_k = -J_\text{H}\sum\limits_{i=1}^3 \big[\Delta\big(\hat{S}_{k,i}^x\hat{S}_{k,i+1}^x+\hat{S}_{k,i}^y\hat{S}_{k,i+1}^y\big)+\hat{S}_{k,i}^z\hat{S}_{k,i+1}^z\big]
       \\     \qquad\quad - J_\text{I}\sum\limits_{i=1}^3 \hat{\sigma}_{k,i}^z \hat{S}_{k,i}^z,
\end{array}
\end{equation}
 where we have imposed periodic boundary conditions by assuming $\hat{S}_{k,4}\equiv\hat{S}_{k,1}$. Introducing the following notation for the 'local fields' $h_{k,1}=J_\text{I}\sigma_{k,1}^z,\;h_{k,2}=J_\text{I}\sigma_{k,2}^z,\;h_{k,3}=J_\text{I}\sigma_{k,3}^z$ allows one to express all eigenvalues of the cluster Hamiltonian (\ref{oneblockH}) under the assumption of the uniform local fields
 $h_{k,1}=h_{k,2}=h_{k,3}=\pm\frac{J_\text{I}}{2}$ 
\begin{equation}\label{eigenenergiesU}
\begin{array}{lcl}
E_{1,2}=-\dfrac{3J_\text{H}}{4}\mp\dfrac{3J_\text{I}}{4},\quad\quad\quad
E_{3,4}=\dfrac{J_\text{H}}{4}\big(1+2\Delta\big)+\dfrac{J_\text{I}}{4},
\\ [0.2cm]
E_{5,6}=\dfrac{J_\text{H}}{4}\big(1+2\Delta\big)-\dfrac{J_\text{I}}{4},\quad
E_{7,8}=\dfrac{J_\text{H}}{4}\big(1-4\Delta\big)\mp\dfrac{J_\text{I}}{4}, 
\end{array}
\end{equation}
as well as non-uniform local fields $h_{k,1}=h_{k,2}=-h_{k,3}=\pm\frac{J_\text{I}}{2}$ (or any other permutation of it)
\begin{equation}\label{eigenenergiesNU}
\begin{array}{lcl}
\Bar{E}_{1,2}=-\dfrac{3J_\text{H}}{4}\mp\dfrac{J_\text{I}}{4},\quad\quad\quad\quad
\Bar{E}_{3,4}=\dfrac{J_\text{H}}{4}\big(1+2\Delta\big)\mp\dfrac{J_\text{I}}{4},\\ [0.25cm]
\Bar{E}_{5,6}=\dfrac{J_\text{H}\big(1-\Delta\big)-J_\text{I}}{4}\mp\dfrac{1}{2}\sqrt{\left(\dfrac{J_\text{H}\Delta}{2}-J_\text{I}\right)^{\!2}
\!\!\!+2\big(J_\text{H}\Delta\big)^2}\!\!,\\[0.25cm]
\Bar{E}_{7,8}=\dfrac{J_\text{H}\big(1-\Delta\big)+J_\text{I}}{4}\mp\dfrac{1}{2}\sqrt{\left(\dfrac{J_\text{H}\Delta}{2}+J_\text{I}\right)^{\!2}\!\!\!+2\big(J_\text{H}\Delta\big)^2}\!\!.
\end{array}
\end{equation}
The partition function of the spin-1/2 Ising-Heisenberg models on two related martini-type lattices can be represented in the following form
\begin{equation}
\begin{array}{lcl}
  Z =\sum\limits_{\{\sigma_i\}}\prod\limits_{k=1}^{\gamma N} \text{Tr}_k\exp\big(-\beta\hat{\mathcal{H}}_k\big)=
  \sum\limits_{\{\sigma_i\}}\prod\limits_{k=1}^{\gamma N} Z_k.
\end{array}
\end{equation}
Here, $\beta=1/(k_{\rm B}T)$, $k_{\rm B}$ is Boltzmann's constant and $T$ is the absolute temperature, the summation over available states of all Ising spins is labeled by the summation symbol $\sum_{\{\sigma_i\}}$, while $\text{Tr}_k$ denotes a trace over spin degrees of freedom of the $k$-th Heisenberg trimer (triangle). After tracing over spin degrees of freedom of the Heisenberg spins, the cluster partition function $Z_k$ can be replaced with the generalized star-triangle transformation \cite{fis59,roj09,str10}
\begin{equation}\label{Z_k}
\begin{array}{lcl}
  Z_k\big(\sigma_{k1}^z,\sigma_{k2}^z,\sigma_{k3}^z\big) =\text{Tr}_k\exp(-\beta \hat{\mathcal{H}}_k)
  \\[0.25cm]
             =A\exp\big[\beta J_\text{eff}\big(\sigma_{k1}^z\sigma_{k2}^z+\sigma_{k2}^z\sigma_{k3}^z+\sigma_{k3}^z\sigma_{k1}^z\big)\big].
\end{array}
\end{equation}
From the mapping relation (\ref{Z_k}) one may derive just two independent equations: one for the uniform spin configurations with three equally aligned Ising spins $h_{k,1}=h_{k,2}=h_{k,3}=\pm\frac{J_\text{I}}{2}$
\begin{equation}\label{TIT_V1}
\begin{array}{lcl}
  V_1\equiv Z_k\big(\mp\frac{1}{2},\mp\frac{1}{2},\mp\frac{1}{2}\big) \\[0.15cm]
 \;\quad = 2\;\exp\big(\frac{3}{4}\beta J_\text{H}\big)\cosh\big(\frac{3}{4}\beta J_\text{I}\big)
 \\[0.15cm]
 \;\quad +4\;\exp\big[-\frac{\beta J_\text{H}}{4}(1+2\Delta)\big]\cosh\big(\frac{\beta J_\text{I}}{4}\big)
  \\[0.15cm]
 \; \quad+2\;\exp\big[-\frac{\beta J_\text{H}}{4}(1-4\Delta)\big]
 \; \cosh\big(\frac{\beta J_\text{I}}{4}\big)
  \\[0.15cm]
\;   \quad = A\exp\big(\frac{3}{4}\beta J_\text{eff}\big),\\
\end{array}
\end{equation}
and other one for six nonuniform spin arrangements $h_{k,1}=h_{k,2}=-h_{k,3}=\pm\frac{J_\text{I}}{2}$ (and all its cyclic permutations) with one Ising spin pointing in opposite with respect to other two \begin{equation}\label{TIT_V2}
\begin{array}{lcl}
    V_2\equiv Z_k\big(\pm\frac{1}{2},\pm\frac{1}{2},\mp\frac{1}{2}\big) =
                   Z_k\big(\pm\frac{1}{2},\mp\frac{1}{2},\pm\frac{1}{2}\big) \\[0.15cm]
                     \quad\; =Z_k\big(\mp\frac{1}{2},\pm\frac{1}{2},\pm\frac{1}{2}\big) \\[0.15cm]
                     \quad\; =2\exp\big({\frac{3}{4}\beta J_\text{H}}\big)\cosh\big(\frac{1}{4}\beta J_\text{I}\big) 
                     \\[0.15cm]
                \;\quad +2 \exp\big({-\frac{\beta J_\text{H}}{4}(1+2\Delta)}\big)\cosh\big(\frac{\beta J_\text{I}}{4}\big)\\[0.15cm]
                \; \quad + 2 \exp\big[{-\frac{\beta}{4}\big( J_\text{H}(1-\Delta)-J_\text{I} \big)}\big]\cosh\big(\frac{\beta}{2}Q_{-}\big) 
 \\[0.15cm]
             \;\quad +\; 2\;\exp\big[{-\frac{\beta}{4}\big(J_\text{H}(1-\Delta)+J_\text{I} \big)}\big]\cosh\big(\frac{\beta}{2}Q_{+}\big) \\[0.15cm]
\quad = A\exp\big(-\frac{1}{4}\beta J_\text{eff}\big).
\end{array}
\end{equation}
For the sake of brevity, we have introduced in above the following notation for two coefficients $Q^{\pm}=\sqrt{\big(\frac{1}{2}J_\text{H}\Delta\pm J_\text{I} \big)^2+2\big( J_\text{H}\Delta \big)^2}$. The mapping parameters $A$ and $J_\text{eff}$ can be straightforwardly obtained by eliminating one of them from the couple of Eqs. (\ref{TIT_V1}) and (\ref{TIT_V2}) 
  \begin{equation}\label{AbetaJeff}
\begin{array}{lcl}
A=\big(V_1V_2^3\big)^{\frac{1}{4}},\quad \beta J_\text{eff}=\ln\Big(\dfrac{V_1}{V_2} \Big).
\end{array}
\end{equation}
Bearing this in mind, the partition function of the spin-1/2 Ising-Heisenberg model on the martini and martini-diced lattices can be obtained from a rigorous mapping relationship to the spin-1/2 Ising model on a simple triangular lattice 
 \begin{equation}
  Z=A^{\gamma N}Z_\text{IM}\big(\beta , \gamma J_\text{eff} \big).
	\label{pf}
\end{equation}
Here, $Z_\text{IM}$ denotes the partition function of the corresponding spin-1/2 Ising model on a triangular lattice given by Hamiltonian $\mathcal{H}_\text{IM}$ 
 \begin{equation}
  \mathcal{H}_\text{IM}=-\gamma J_\text{eff}\sum\limits_{\langle i,j\rangle}\hat{\sigma}_i^z\hat{\sigma}_j^z,
\end{equation}
which is expressed only through the effective nearest-neighbor interaction $\gamma J_\text{eff}$. The size of the effective nearest-neighbor interaction $\gamma J_\text{eff}$ is in fact the most important characteristics, which distinguishes the magnetic behavior of both lattice-statistical models. To be more specific, the size of effective coupling $\beta\gamma J_\text{eff}$ for the spin-1/2 Ising-Heisenberg martini-diced lattice shown in Fig. \ref{fig:TIT3}(b) is twice as strong as that for the martini lattice shown in Fig. \ref{fig:TIT3}(a) due to two times higher number of local star-triangle transformations. Finally, let us also quote the exact result for the partition function of the spin-1/2 Ising model on a triangular lattice \cite{wan50,hou50,new50} that completes our exact calculation of the partition function
\begin{equation}
\begin{array}{lcl}
  \ln({{Z_\text{IM}}/{2}})=\dfrac{1}{8\pi^2}\mathlarger{\int}_0^{2\pi}\mathlarger{\int}_0^{2\pi}\ln\big[C_1-D_1\big(\cos(\theta)+\cos(\phi)\\
   \qquad\qquad  \qquad\qquad  \qquad\qquad\quad + \cos(\theta+\phi)\big)\big]d\theta d\phi
 \end{array}
\end{equation}
where
\begin{equation}
\begin{array}{lcl}
    C_1=\cosh^3\Big(\dfrac{\beta\gamma J_\text{eff}}{2}\Big)+\sinh^3\Big(\dfrac{\beta\gamma J_\text{eff}}{2}\Big),\\[0.25cm]
     D_1=\sinh\Big(\dfrac{\beta\gamma J_\text{eff}}{2}\Big).
\end{array}
\end{equation}  

Other thermodynamic quantities can be now straightforwardly derived from the mapping relation (\ref{pf}) between the partition functions. For example, the free energy of the spin-1/2 Ising-Heisenberg model on the martini and martini-diced lattices can be written as  $F=F_\text{IM}(\beta,\gamma J_\text{eff})-\gamma Nk_BT\ln A$, whereby $F_\text{IM}=-k_BT\ln Z_\text{IM}$ represents the free energy of
the spin-1/2 Ising model on the triangular lattice \cite{wan50,hou50,new50}. The internal energy $U$ of the spin-1/2 Ising-Heisenberg model on the martini and martini-diced lattices can be also linked to the internal energy $U_\text{IM}$ of the effective spin-1/2 Ising model on the triangular lattice via the following mapping relation
\begin{equation}\label{TIT_U}
\begin{array}{lcl}
\dfrac{U}{N}=-\dfrac{\gamma}{4}\Big( \dfrac{W_1}{V_1}+ 3\dfrac{W_2}{V_2} \Big)+\dfrac{U_\text{IM}}{NJ_\text{eff}}\Big( \dfrac{W_1}{V_1}- \dfrac{W_2}{V_2} \Big),
\end{array}
\end{equation}
where two new functions $W_1=\partial V_1/\partial \beta$ and $W_2=\partial V_2/\partial \beta$ can be calculated from Eqs. (\ref{TIT_V1}) and (\ref{TIT_V2}) by performing the respective derivatives with respect to an inverse temperature $\beta$. Note furthermore that the exact result for the internal energy $U_\text{IM}$ of the spin-1/2 Ising model on the triangular lattice can be for instance found in Refs. \cite{wan50,dom60}. Last but not least, the specific heat of the investigated models can be acquired from the temperature derivative of the internal energy $C=\partial U/\partial T$.

The spontaneous magnetization of the Ising spins $m_\text{I}\equiv \langle\hat{\sigma}_i^z\rangle$ of the spin-1/2 Ising-Heisenberg model on the martini and martini-diced lattices relates to the spontaneous magnetization $m_\text{IM}\equiv\langle\hat{\sigma}_i^z\rangle_\text{IM}$ of the effective spin-1/2 Ising model on the triangular lattice \cite{pot52} on account of the equality of both sublattice magnetizations $m_\text{I}=m_\text{IM}(\beta, \gamma J_\text{eff})$. For the sake of completeness, let us therefore quote also the exact result for the spontaneous magnetization of the effective spin-1/2 Ising model on a triangular lattice \cite{pot52} 
 \begin{equation}\label{TIT_I}
\begin{array}{lcl}
m_\text{IM}=\dfrac{1}{2}\Bigg[1- \dfrac{16 z^6}{\big(1+3 z^2\big)\big( 1- z^2  \big)^3}\Bigg]^{\frac{1}{8}},
\end{array}
\end{equation}
which is expressed in terms of the parameter $z=\exp(-\gamma\beta J_\text{eff}/2)$. The spontaneous magnetization of the Ising spins can be thus obtained by substituting the exact result (\ref{AbetaJeff}) for the mapping parameter $\beta J_\text{eff}$ into Eq. (\ref{TIT_I}) 
 \begin{equation}\label{TIT_mI}
\begin{array}{lcl}
m_\text{I}=\dfrac{1}{2}\Bigg[1- \dfrac{16 V_1^{\gamma}V_2^{3\gamma}}{\big(V_1^{\gamma}+3V_2^{\gamma}\big)\big( V_1^{\gamma}- V_2^{\gamma}  \big)^3}\Bigg]^{\frac{1}{8}}.
\end{array}
\end{equation}
On the other hand, the spontaneous magnetization of the Heisenberg spins $m_\text{H} \equiv \langle \hat{S}_{k,i}^z \rangle$ can be relatively straightforwardly evaluated with the help of the generalized Callen-Suzuki spin identity \cite{cal63,suz65,bal02} 
\begin{eqnarray}
m_{\rm H} \equiv \langle \hat{S}_{k,i}^z \rangle =  \left \langle \frac{\mbox{Tr}_k \hat{S}_{k,i}^z \exp(-\beta \hat{\cal H}_k)}{\mbox{Tr}_k \exp(- \beta \hat {\cal H}_k)}\right \rangle,
\label{eq:csu}
\end{eqnarray}
which merely involves a local trace over degrees of freedom of three Heisenberg spins belonging to the $k$th martini unit cell denoted by the symbol $\mbox{Tr}_k$. After performing the relevant trace one obtains the resultant expression, which depends on all possible odd statistical mean values developed from three Ising spins involved in the cluster Hamiltonian (\ref{oneblockH}). The spontaneous magnetization of the Heisenberg spins can be consequently related to the spontaneous magnetization $m_\text{IM}\equiv\langle\hat{\sigma}_i^z\rangle_\text{IM}$ and the triplet (three-spin) correlation function $t_\text{IM}\equiv\langle \hat{\sigma}_{k,1}^z\hat{\sigma}_{k,2}^z\hat{\sigma}_{k,3}^z\rangle_\text{IM}$ of the effective spin-1/2 Ising model on a triangular lattice through the exact mapping relation
\begin{equation}\label{TIT_mH}
\begin{array}{lcl}
m_\text{H}=\dfrac{m_\text{IM}}{2}\Big(\dfrac{Q_1}{V_1}+\dfrac{Q_2}{V_2}\Big)+\dfrac{2t_\text{IM}}{3}\Big(\dfrac{Q_1}{V_1}-3\dfrac{Q_2}{V_2}\Big),
\end{array}
\end{equation}
where two functions $Q_1$ and $Q_2$ are defined as
\begin{equation}\label{Q1Q2}
\begin{array}{lcl}
Q_1 = {3}\exp\Big(\dfrac{3}{4}\beta J_\text{H}\Big)\sinh\Big(\dfrac{3}{4}\beta J_\text{I}\Big)\\[0.2cm]
+\exp\Big[-\dfrac{1}{4}\beta J_\text{H}(1-4\Delta)\Big]\sinh\Big(\dfrac{1}{4}\beta J_\text{I}\Big)\\[0.2cm]
+2\exp\Big[-\dfrac{1}{4}\beta J_\text{H}(1+2\Delta)\Big]\sinh\Big(\dfrac{1}{4}\beta J_\text{I}\Big),\\[0.3cm]
Q_2 = {3}\exp\Big(\dfrac{3}{4}\beta J_\text{H}\Big)\sinh\Big(\dfrac{3}{4}\beta J_\text{I}\Big)\\[0.2cm]
-\exp\Big[-\dfrac{1}{4}\beta J_\text{H}(1+2\Delta)\Big]\sinh\Big(\dfrac{1}{4}\beta J_\text{I}\Big)\\[0.2cm]
+\exp\Big[-\dfrac{1}{4}\beta (J_\text{H}(1-\Delta)-J_\text{I})\Big]\cosh\Big(\dfrac{1}{2}\beta Q^{-}\Big),\\[0.2cm]
-\exp\Big[-\dfrac{1}{4}\beta (J_\text{H}(1-\Delta)+J_\text{I})\Big]\cosh\Big(\dfrac{1}{2}\beta Q^{+}\Big).\\[0.3cm]
\end{array}
\end{equation}
It is evident from Eq. (\ref{TIT_mH}) that the spontaneous magnetization of the Heisenberg spins $m_\text{H}$ can be expressed in terms of the formerly derived the spontaneous magnetization of the Ising spins $m_\text{IM}$, while the three-spin correlation function $t_\text{IM}$ follows from a rigorous calculation by Baxter and Choy \cite{bax89} 
\begin{equation}\label{t_IM}
\begin{array}{lcl}
t_\text{IM}=\dfrac{m_\text{IM}}{4}\Bigg[1+ 2\dfrac{V_1^{2\gamma}-2V_2^{2\gamma}+V_1^{\gamma}V_2^{\gamma}}{\big(V_1^{\gamma}-V_2^{\gamma}\big)^2} - \\[0.5cm]
\qquad\qquad \qquad \dfrac{2V_1^{\gamma}
\sqrt{\big(V_1^{\gamma}+3V_2^{\gamma}\big)\big( V_1^{\gamma}- V_2^{\gamma}  \big)}}{\big(V_1^{\gamma}-V_2^{\gamma}\big)^2}
\Bigg].
\end{array}
\end{equation}

\section{Results and discussion}\label{results}
In this section, we will proceed to a detailed examination of the ground-state and finite-temperature phase diagrams, spontaneous magnetization and thermodynamic properties of the spin-1/2 Ising-Heisenberg model on the martini and martini-diced lattices. In what follows the size of the ferromagnetic Ising coupling constant $J_\text{I}>0$ will be used as an energy unit when defining two reduced parameters: the dimensionless temperature $k_B T/ J_\text{I}$ and the relative strength of the Heisenberg interaction with respect to the Ising coupling constant $J_\text{H}/J_\text{I}$. It is noteworthy that the magnetic behavior of the spin-1/2 Ising-Heisenberg model on the martini and martini-diced lattices with the antiferromagnetic Ising coupling constant $J_\text{I}<0$ is completely identical except a trivial flip of all the Ising spins, so this particular case will be left out from our consideration for simplicity. 

\begin{figure}
\begin{center}
\resizebox{0.5\textwidth}{!}{\includegraphics[clip]{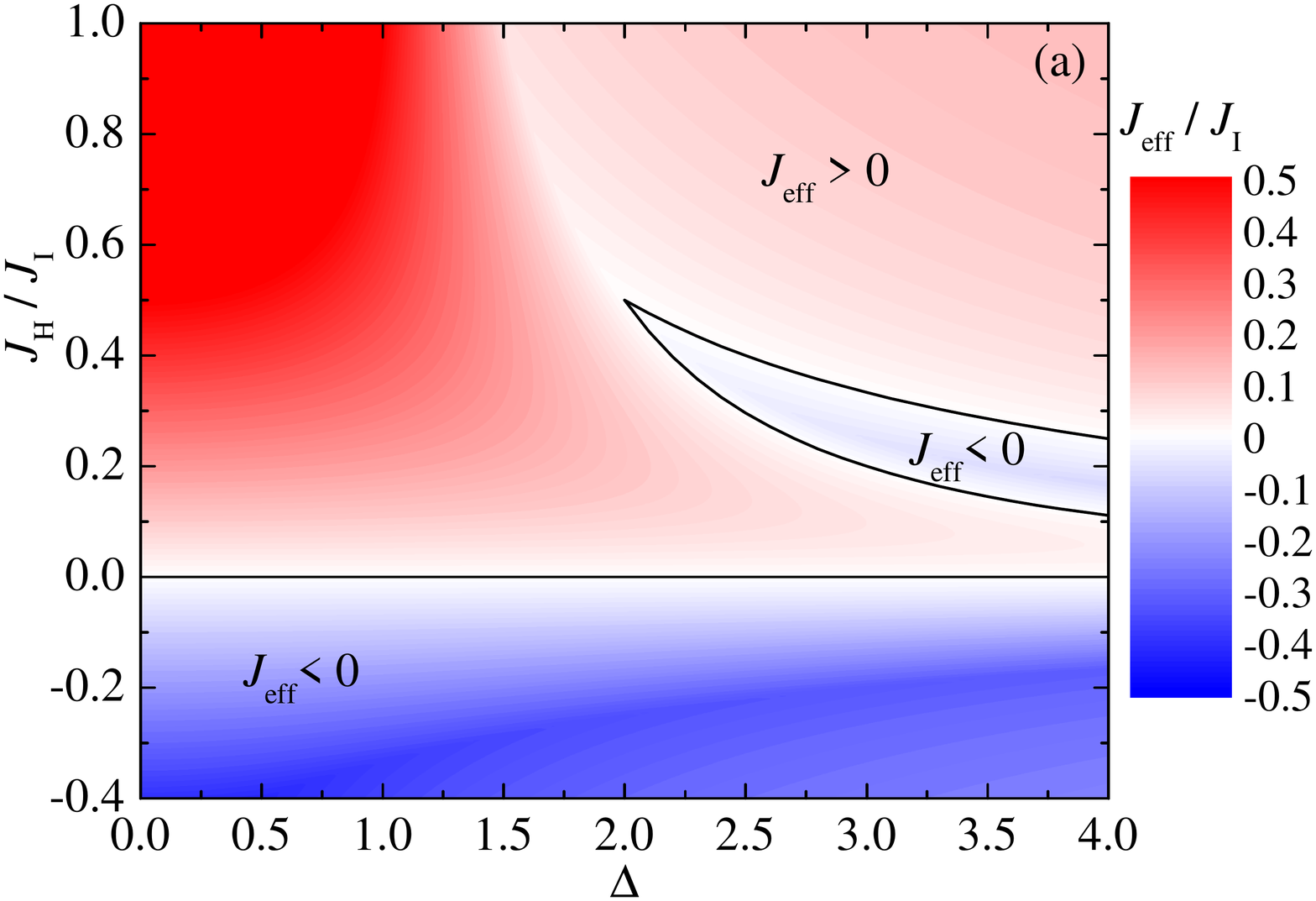}}
\resizebox{0.5\textwidth}{!}{\includegraphics[clip]{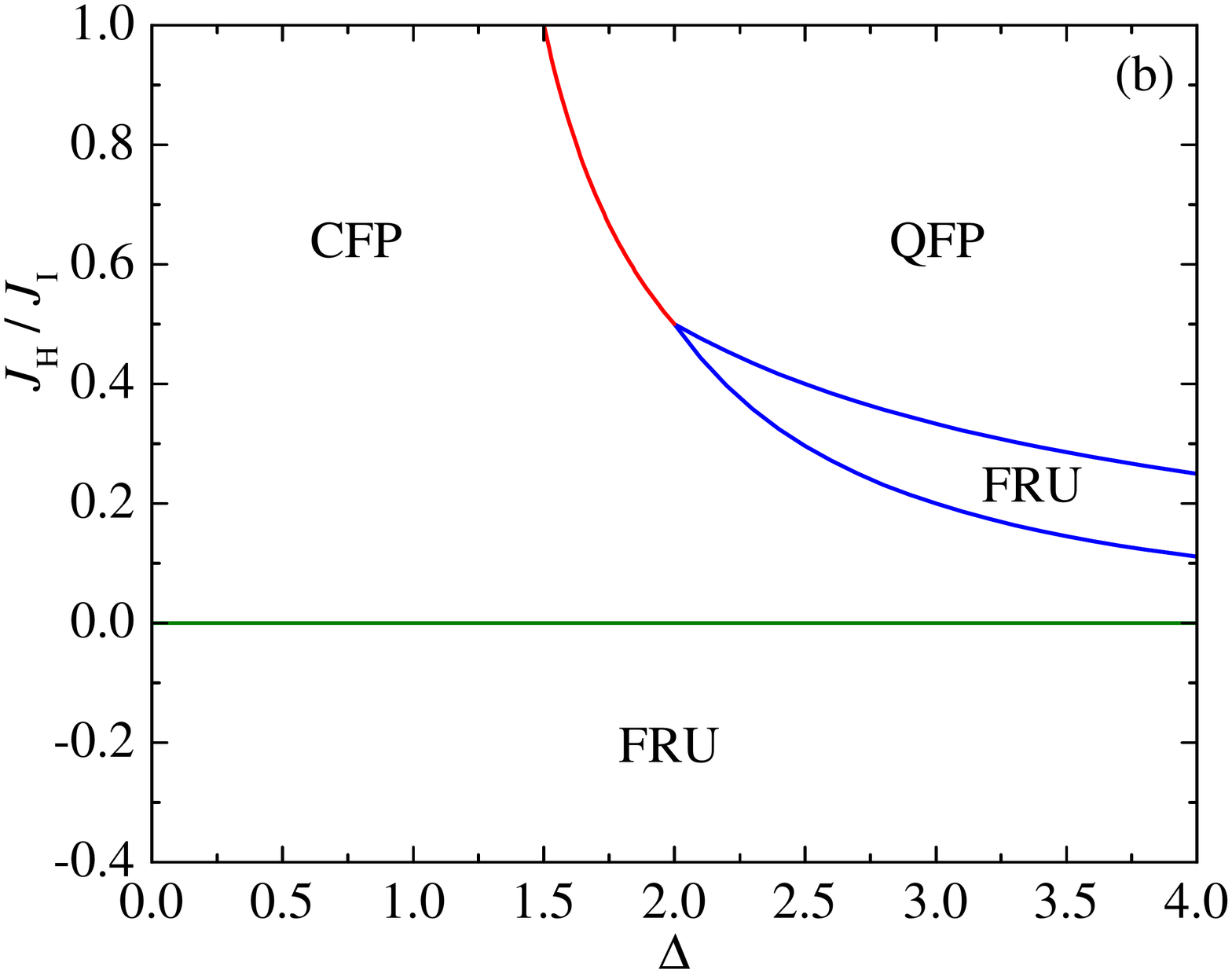}}
\vspace*{-1.3cm}
\caption{(a) A density plot of zero-temperature asymptotic value of the effective coupling $J_\text{eff}$ given by Eq. (\ref{AbetaJeff}) in the $\Delta-J_\text{H}/J_\text{I}$ plane. The effective interaction $J_\text{eff}$ changes its sign along displayed black contour lines; (b) The ground-state phase diagram of the spin-1/2 Ising-Heisenberg model on the martini and martini-diced lattices in the $\Delta-J_\text{H}/J_\text{I}$ plane.}
\label{fig:QFT_TIT}       
\end{center}
\end{figure}

\subsection{Effective interaction at zero temperature}
In the previous section we have proved that the spin-1/2 Ising-Heisenberg model on the martini and martini-diced lattices can be exactly mapped to the spin-1/2 Ising model on a triangular lattice with the effective temperature-dependent pair interaction $\gamma J_\text{eff}$ given by Eq. (\ref{AbetaJeff}). It is well established that the spin-1/2 Ising model on a triangular lattice with the ferromagnetic pair interaction ($J_\text{eff}>0$) exhibits at low enough temperatures a spontaneous long-range order, whereas a spontaneous ordering is totally absent at any finite temperature for the analogous model with antiferromagnetic pair interaction ($J_\text{eff}<0$) \cite{wan50,hou50,new50}. Owing to this fact, a sign of zero-temperature asymptotic limit of the effective interaction $J_\text{eff}$ decisively determines whether the spin-1/2 Ising-Heisenberg model on the martini and martini-diced lattices shows a spontaneous order (for $J_\text{eff}>0$) or disorder (for $J_\text{eff}<0$). A zero-temperature density plot of the effective interaction, which is displayed in Fig. \ref{fig:QFT_TIT}(a) including zero contour lines $J_\text{eff} (T=0) = 0$, accordingly delimits respective parameter regions pertinent to a spontaneous order and disorder. Although presence of the disordered ground state corresponding to the effective antiferromagnetic interaction $J_\text{eff}<0$ is somewhat expected for the antiferromagnetic Heisenberg coupling constant $J_\text{H}<0$ promoting a geometric spin frustration, the other disordered ground state unexpectedly appears in a relatively narrow parameter region $J_\text{H}/J_\text{I} \in (\frac{2}{(\Delta+2)(\Delta-1)}, \frac{1}{\Delta})$ with purely ferromagnetic interactions $J_\text{I}>0$ and $J_\text{H}>0$ under the assumption of a sufficiently strong easy-plane exchange anisotropy $\Delta>2$. In addition, the abrupt change in a size of the effective coupling observed in the parameter space with $J_\text{eff}>0$ indicates a possible existence of two different spontaneously ordered phases. 

\subsection{Ground-state phase diagram}
Let us bring insight into a microscopic nature of individual ground states of the spin-1/2 Ising-Heisenberg model on the martini and martini-diced lattices, which can be found from lowest-energy eigenstates of the cluster Hamiltonian (\ref{oneblockH}) inherent to a single martini-type six-spin cluster serving as a basic building block of both martini-type lattices. The spin-1/2 Ising-Heisenberg model on two related lattices accordingly have identical ground states and zero-temperature phase boundaries, which can be all identified as the lines of discontinuous (first-order) phase transitions. The ground-state phase diagram in the $\Delta-J_\text{H}/J_\text{I}$ plane is quite reminiscent of the density plot of the effective interaction [c.f. Fig. \ref{fig:QFT_TIT}(b) with Fig. \ref{fig:QFT_TIT}(a)], whereby it involves two spontaneously long-range ordered and one disordered ground state referred to as the classical ferromagnetic phase (CFP), the quantum ferromagnetic phase (QFP) and the frustrated phase (FRU). Phase boundaries between these ground states are given by following conditions: 
 \begin{equation}\label{GSPT}
\begin{array}{lcl}
\text{QFP}-\text{CFP}: J_\text{H}/J_\text{I}=\dfrac{1}{2(\Delta-1)}, \quad (1\leqq\Delta\leqq 2)\\[0.4cm]
\text{CFP}-\text{FRU}:
\left\{
  \begin{array}{@{}l@{}}
J_\text{H}/J_\text{I}=\dfrac{2}{(\Delta+2)(\Delta-1)}, \quad (\Delta\geqq 2)\\[0.4cm]
 J_\text{H}/J_\text{I}=0
  \end{array}\right. \\[0.6cm]
\text{QFP}-\text{FRU}: J_\text{H}/J_\text{I}=\dfrac{1}{\Delta}, \quad  (\Delta\geqq 2).
\end{array}
\end{equation}
The first spontaneously ordered ground state CFP with energy $E_1$ per martini unit cell is two-fold degenerate and is given by the eigenvectors
 \begin{equation}\label{CFP}
\begin{array}{lcl}
\vert \text{CFP} \rangle =
\left\{
  \begin{array}{@{}l@{}}
 \prod\limits_{i=1}^N\vert\!\uparrow\rangle_{\sigma_i^z}\otimes\prod\limits_{k=1}^{\gamma N}\vert\!\uparrow\uparrow\uparrow\rangle_{S_{k1}^z,S_{k2}^z,S_{k3}^z},
 \\[0.3cm]
  \prod\limits_{i=1}^N\vert\!\downarrow\rangle_{\sigma_i^z}\otimes\prod\limits_{k=1}^{\gamma N}\vert\!\downarrow\downarrow\downarrow\rangle_{S_{k1}^z,S_{k2}^z,S_{k3}^z}.
 \end{array}\right.
\end{array}
\end{equation}
The second spontaneously ordered ground state QFP with energy $E_7$ per martini unit cell is likewise two-fold degenerate and is unambiguously given by the eigenvectors
\begin{equation}\label{QFP}
  \begin{array}{@{}l@{}}
\vert \text{QFP} \rangle = \\[0.25cm]
\left\{
  \begin{array}{@{}l@{}}
\prod\limits_{i=1}^N\vert\!\uparrow\rangle_{\sigma_i^z}\otimes
\prod\limits_{k=1}^{\gamma N}\dfrac{1}{\sqrt{3}}(\vert\!\uparrow\uparrow\downarrow\rangle+\vert\!\uparrow\downarrow\uparrow\rangle+\vert\!\downarrow\uparrow\uparrow\rangle)_{S_{k1}^z,S_{k2}^z,S_{k3}^z},
\\[0.3cm]
\prod\limits_{i=1}^N\!\vert\downarrow\rangle_{\sigma_i^z}\otimes
\prod\limits_{k=1}^{\gamma N}\dfrac{1}{\sqrt{3}}(\vert\!\downarrow\downarrow\uparrow\rangle+\vert\downarrow\uparrow\downarrow\rangle+\vert\!\uparrow\downarrow\downarrow\rangle)_{S_{k1}^z,S_{k2}^z,S_{k3}^z}.
\end{array}\right.
\end{array}
\end{equation}
While the former ground state CFP (\ref{CFP}) shows a classical ferromagnetic long-range order with fully saturated magnetic moments of all Ising as well as Heisenberg spins, the latter ground state QFP (\ref{QFP}) is obviously of purely quantum nature owing to a symmetric quantum superposition of three up-up-down (uud) or down-down-up (ddu) microstates of three Heisenberg spins. 

However, the most remarkable quantum spin arrangement is found within the disordered ground state FRU with energy $\Bar{E}_5$ per martini unit cell, which refers to a macroscopically degenerate manifold of eigenstates constructed from the following set of six energetically equivalent uud and ddu states of three Ising and Heisenberg spins belonging to the same martini unit cell
\begin{equation}\label{QPP}
\begin{array}{lcl}
\vert \text{uud}\rangle = \\[0.25cm]
\left\{
  \begin{array}{@{}l@{}}
\vert\!\!\uparrow\uparrow\downarrow\rangle_{\sigma_1^z,\sigma_2^z,\sigma_3^z}\!\otimes\!\! \left[\sin \! \phi \vert\!\!\uparrow\uparrow\downarrow\rangle \!+\! \frac{\cos \! \phi}{\sqrt{2}} \! \left(\vert\!\!\uparrow\downarrow\uparrow\rangle \!\!+\!\! \vert\!\!\downarrow\uparrow\uparrow\rangle \right)\right]_{\!S_{k1}^z,S_{k2}^z,S_{k3}^z} \\
\vert\!\!\uparrow\downarrow\uparrow\rangle_{\sigma_1^z,\sigma_2^z,\sigma_3^z}\!\otimes\!\! \left[\sin \! \phi \vert\!\!\uparrow\downarrow\uparrow\rangle \!+\! \frac{\cos \! \phi}{\sqrt{2}} \! \left(\vert\!\!\uparrow\uparrow\downarrow\rangle \!\!+\!\! \vert\!\!\downarrow\uparrow\uparrow\rangle \right)\right]_{\!S_{k1}^z,S_{k2}^z,S_{k3}^z} \\
\vert\!\!\downarrow\uparrow\uparrow\rangle_{\sigma_1^z,\sigma_2^z,\sigma_3^z}\!\otimes\!\! \left[\sin \! \phi \vert\!\!\downarrow\uparrow\uparrow\rangle \!+\! \frac{\cos \! \phi}{\sqrt{2}} \! \left(\vert\!\!\uparrow\uparrow\downarrow\rangle \!\!+\!\! \vert\!\!\uparrow\downarrow\uparrow\rangle \right)\right]_{\!S_{k1}^z,S_{k2}^z,S_{k3}^z} \\
\end{array}\right. 
\\[0.75cm]
\vert \text{ddu} \rangle = \\[0.25cm]
\left\{
  \begin{array}{@{}l@{}}
\vert\!\!\downarrow\downarrow\uparrow\rangle_{\sigma_1^z,\sigma_2^z,\sigma_3^z}\!\otimes\!\! \left[\sin \! \phi \vert\!\!\downarrow\downarrow\uparrow\rangle \!+\! \frac{\cos \! \phi}{\sqrt{2}} \! \left(\vert\!\!\downarrow\uparrow\downarrow\rangle 
\!\!+\!\! \vert\!\!\uparrow\downarrow\downarrow\rangle \right)\right]_{\!S_{k1}^z,S_{k2}^z,S_{k3}^z} \\
\vert\!\!\downarrow\uparrow\downarrow\rangle_{\sigma_1^z,\sigma_2^z,\sigma_3^z}\!\otimes\!\! \left[\sin \! \phi \vert\!\!\downarrow\uparrow\downarrow\rangle \!+\! \frac{\cos \! \phi}{\sqrt{2}} \! \left(\vert\!\!\downarrow\downarrow\uparrow\rangle \!\!+\!\! \vert\!\!\uparrow\downarrow\downarrow\rangle \right)\right]_{\!S_{k1}^z,S_{k2}^z,S_{k3}^z} \\
\vert\!\!\uparrow\downarrow\downarrow\rangle_{\sigma_1^z,\sigma_2^z,\sigma_3^z}\!\otimes\!\! \left[\sin \! \phi \vert\!\!\uparrow\downarrow\downarrow\rangle \!+\! \frac{\cos \! \phi}{\sqrt{2}} \! \left(\vert\!\!\downarrow\downarrow\uparrow\rangle \!\!+\!\! \vert\!\!\downarrow\uparrow\downarrow\rangle \right)\right]_{\!S_{k1}^z,S_{k2}^z,S_{k3}^z} \\
\end{array}\right. \\[0.5cm]
\end{array}
\end{equation}
where the mixing angle $\phi$ determining the relevant probability amplitudes is given by $\phi = \frac{1}{2} \arctan [\sqrt{8} J_\text{H} \Delta/(J_\text{H} \Delta - 2 J_\text{I})]$. It should be pointed out, moreover, that the disordered ground state FRU (\ref{QPP}) may emerge not only due to a spin frustration caused by the antiferromagnetic Heisenberg interaction $J_\text{H}<0$, but it may more strikingly appear also in the unfrustrated parameter space where 'unconventional spin frustration' results from a mutual competition between two different ferromagnetic interactions $J_\text{I}>0$ and $J_\text{H}>0$ of easy-axis and easy-plane type, respectively. It is noteworthy that the ground-state spin arrangement of the Heisenberg trimers (triangles) is within the ground state FRU (\ref{QPP}) unique, since it is unambiguously determined by the relevant spin configuration of three enclosing Ising spins. Hence, it follows that the residual entropy per spin $S_{\rm res}/[N(1+3\gamma) k_B]$ ascribed to the disordered ground state FRU (\ref{QPP}) of the spin-1/2 Ising-Heisenberg model on the martini and martini-diced lattices can be readily related to the residual entropy of the effective spin-1/2 Ising model on the triangular lattice \cite{wan50} when performing the relevant conversion with respect to the total number of spins, i.e. the residual entropy per spin of the martini and martini-diced lattice accordingly equals to $S_{\rm res}/(4Nk_B) \approx  0.08077$ and $S_{\rm res}/(7Nk_B) \approx 0.04615$, respectively. It also apparent from Fig. \ref{fig:QFT_TIT} that all three ground states CFP, QFP and FRU coexist together at a triple point with the coordinates $[\Delta,J_\text{H}/J_\text{I}]=[2,\frac{1}{2}]$.

\begin{figure*}
\begin{center}
\includegraphics[scale=0.4,trim=0   20 0 20, clip]{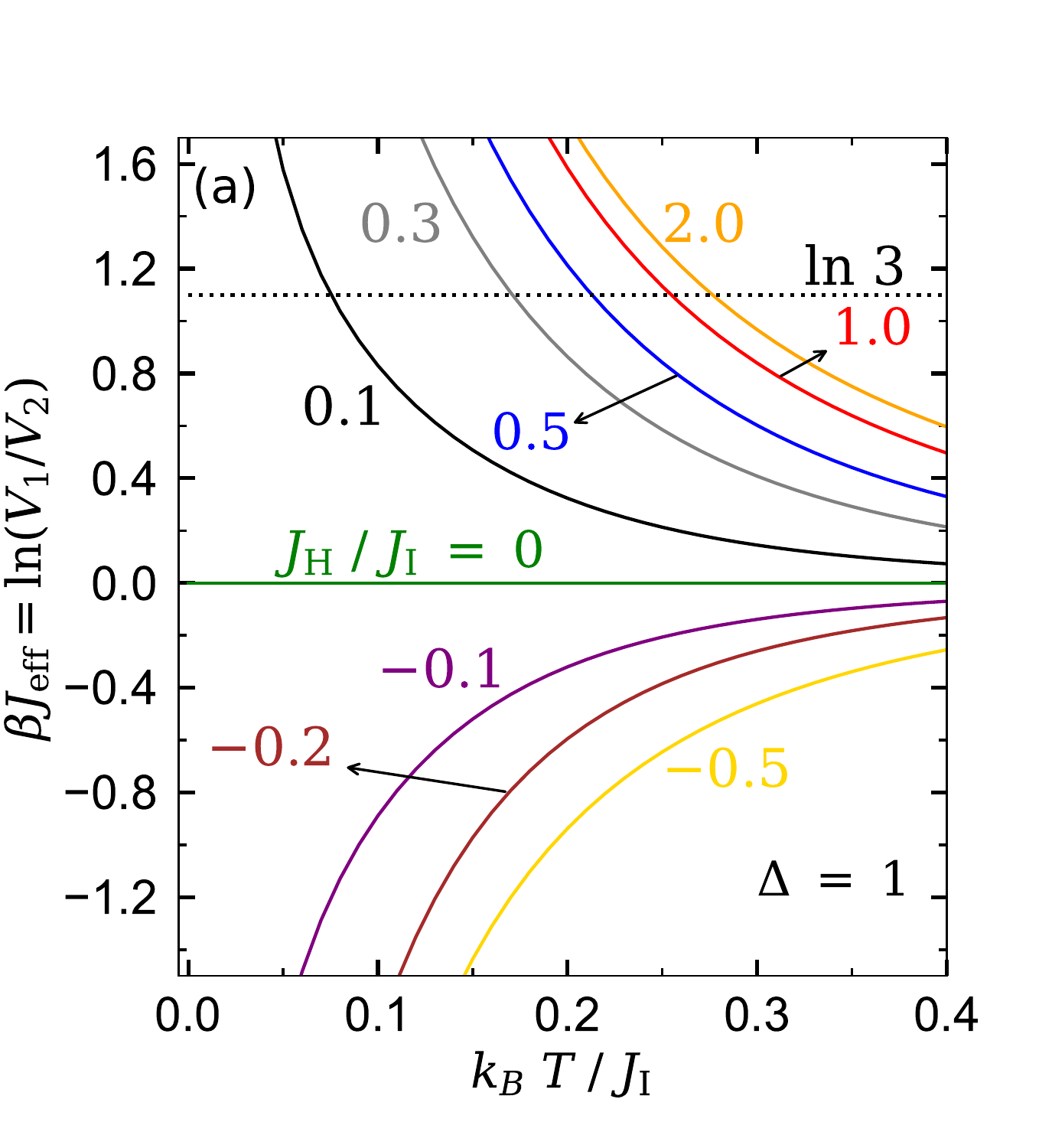}
\includegraphics[scale=0.4,trim=0 20 0 20, clip]{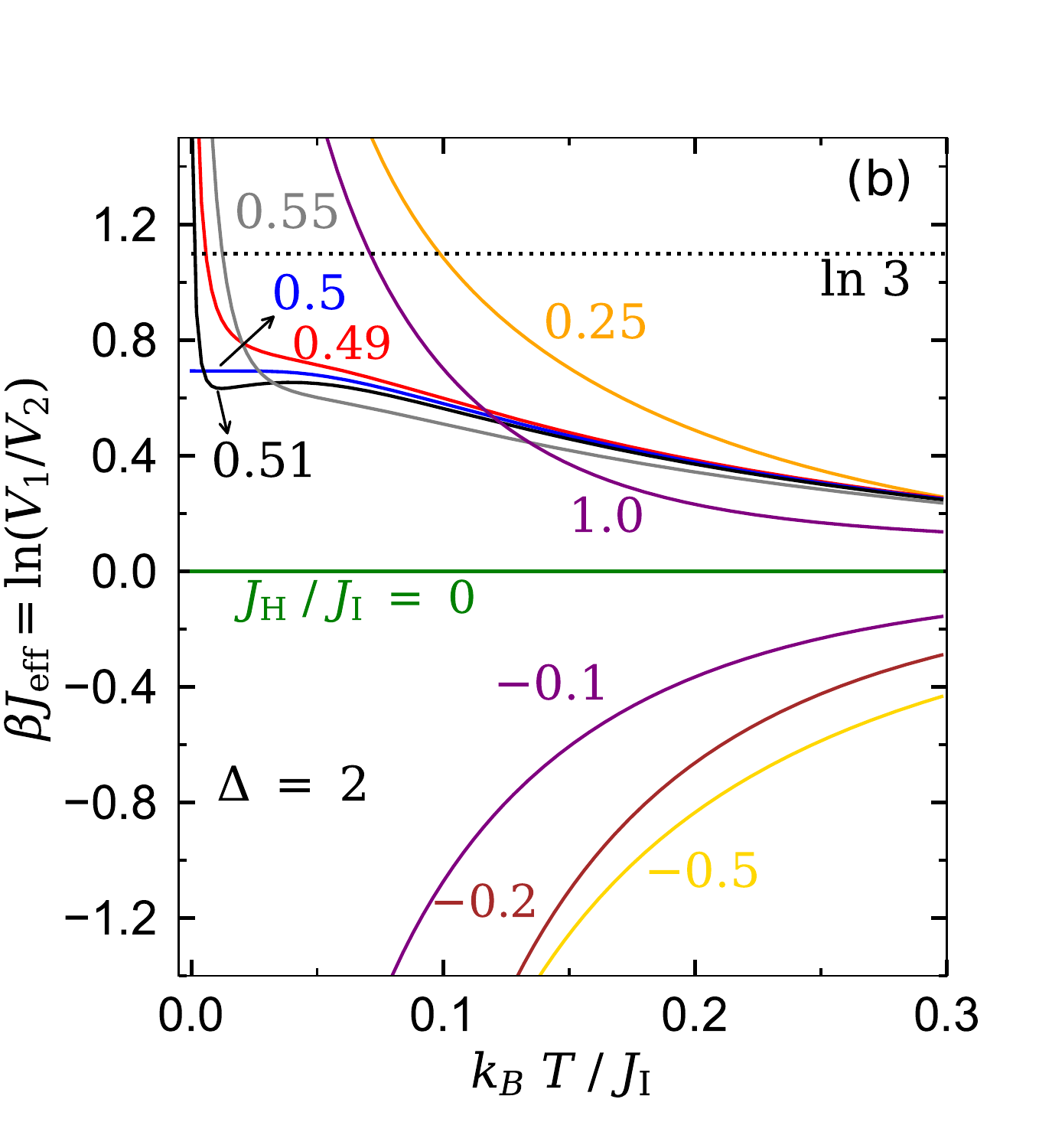}
\includegraphics[scale=0.4,trim= 0  20 0   20, clip]{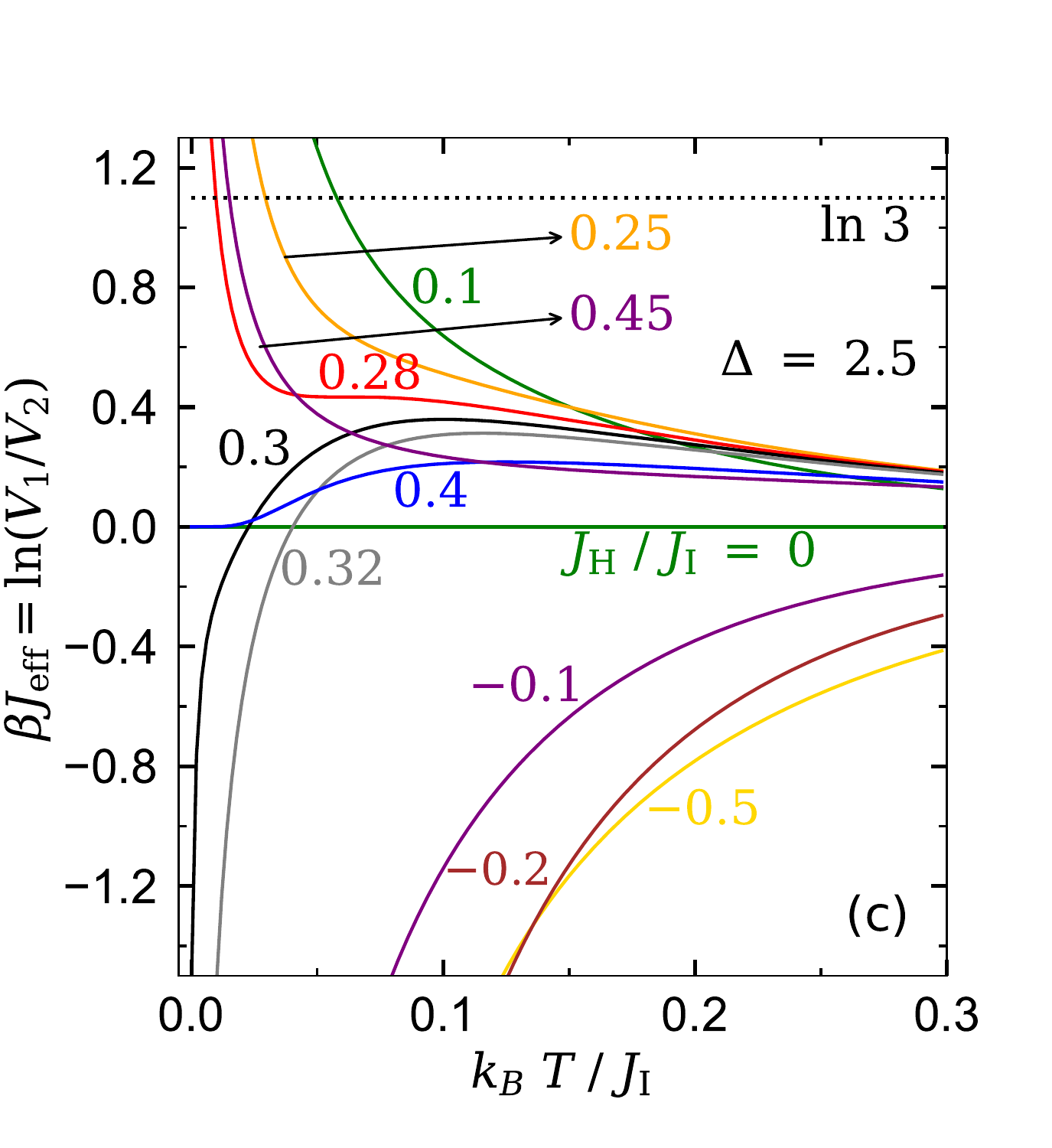}
\caption{The effective coupling $\beta J_\text{eff}$ as a function of the temperature $k_BT/J_\text{I}$ for several values of the interaction ratio $J_\text{H}/J_\text{I}$ and three selected values of the exchange anisotropy: (a) $\Delta=1$, (b) $\Delta=2$ and (c) $\Delta=2.5$. A dotted line shows the critical value of the effective coupling $\beta_c J_\text{eff} = \ln 3$, above which a spontaneous long-range order is present.}
\label{fig:Critical_gamma1}       
\end{center}
\end{figure*}

\subsection{Effective interaction at finite temperatures}
The critical temperature of the spin-1/2 Ising-Heisenberg model on the martini and martini-diced lattices can be readily computed from the critical condition requiring equality of the effective coupling with its critical value 
\begin{equation}\label{CCond}
\begin{array}{lcl}
\gamma \beta_\text{c} J_\text{eff}=\ln \Big(\dfrac{V_1^{\gamma}}{V_2^{\gamma}}  \Big)=
\ln 3 \Longrightarrow V_1^{\gamma}(\beta_\text{c})={3}V_2^{\gamma}(\beta_\text{c}).
\end{array}
\end{equation}
It could be deduced from the critical condition (\ref{CCond}) that the spin-1/2 Ising-Heisenberg model on the martini-type lattices exhibits a spontaneous long-range order if the effective coupling exceeds the critical value $\gamma\beta J_\text{eff} > \ln 3$, while a disordered spin state emerges in the reverse case $\gamma\beta J_\text{eff} < \ln 3$. For better understanding, the effective coupling $\beta J_\text{eff}$ of the spin-1/2 Ising-Heisenberg model on the martini lattice is depicted in Fig. \ref{fig:Critical_gamma1} against temperature for three different values of the exchange anisotropy. It can be easily understood from this figure that the effective coupling diverges to infinity as temperature tends to zero for the isotropic ferromagnetic coupling constant $J_\text{H} >0, \Delta=1$ due to the spontaneously ordered ground state CFP, while it diverges to minus infinity for the isotropic antiferromagnetic coupling constant $J_\text{H} <0, \Delta=1$ due to the disordered ground state FRU [see Fig. \ref{fig:Critical_gamma1}(a)]. Contrary to this, the effective coupling  $\beta J_\text{eff}$ tends towards a constant value at the triple coexistence point $[\Delta, J_\text{H}/J_\text{I}]=[2,\frac{1}{2}]$ of all three ground states CFP, QFP and FRU [see Fig. \ref{fig:Critical_gamma1}(b)]. Moreover, the zero-temperature divergence of the effective coupling at the specific value of the exchange anisotropy $\Delta=2$ is due to the ground state CFP only for $J_\text{H}/J_\text{I}<1/2$, while the effective coupling diverges for $J_\text{H}/J_\text{I}>1/2$ due to the other spontaneously ordered ground state QFP. The most 
diverse behavior of the effective coupling can be however detected by considering sufficiently strong exchange anisotropies $\Delta>2$ as exemplified in Fig. \ref{fig:Critical_gamma1}(c) for the particular case $\Delta=2.5$. Apart from the behavior reported on previously, 
the effective coupling shows in a limited range of the ferromagnetic coupling constant (e.g. $8/27<J_\text{H}/J_\text{I}<2/5$ for 
$\Delta=2.5$) a remarkable nonmonotonous temperature dependence: the effective coupling first steadily rises with decreasing of temperature until it reaches a relatively broad maximum, which is successively followed by a sudden divergence down to minus infinity emergent due to the frustrated ground state FRU.

\begin{figure*}
\begin{center}
\includegraphics[scale=0.5,trim=10 10 20 40, clip]{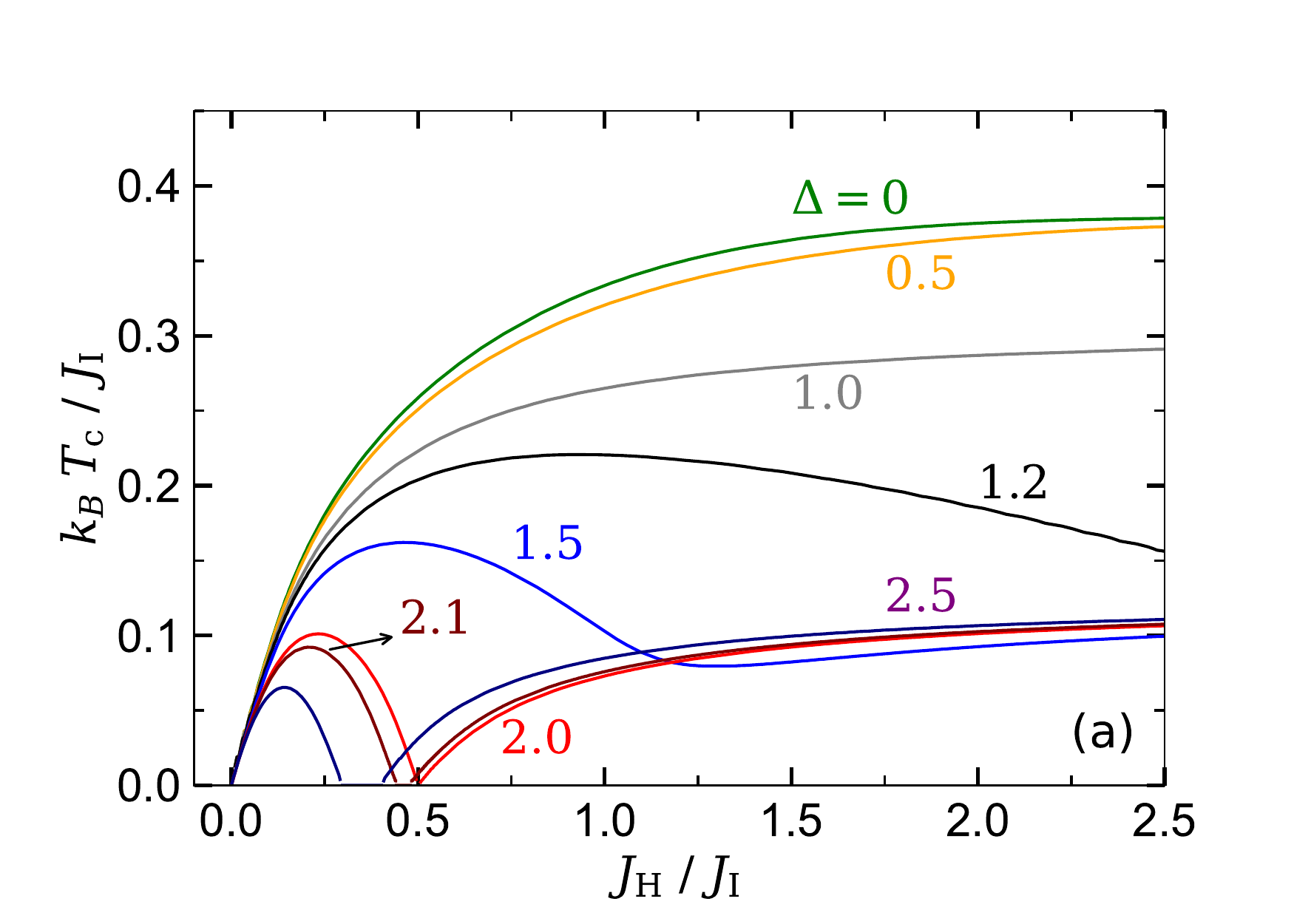}
\includegraphics[scale=0.5,trim=20 10 10 40, clip]{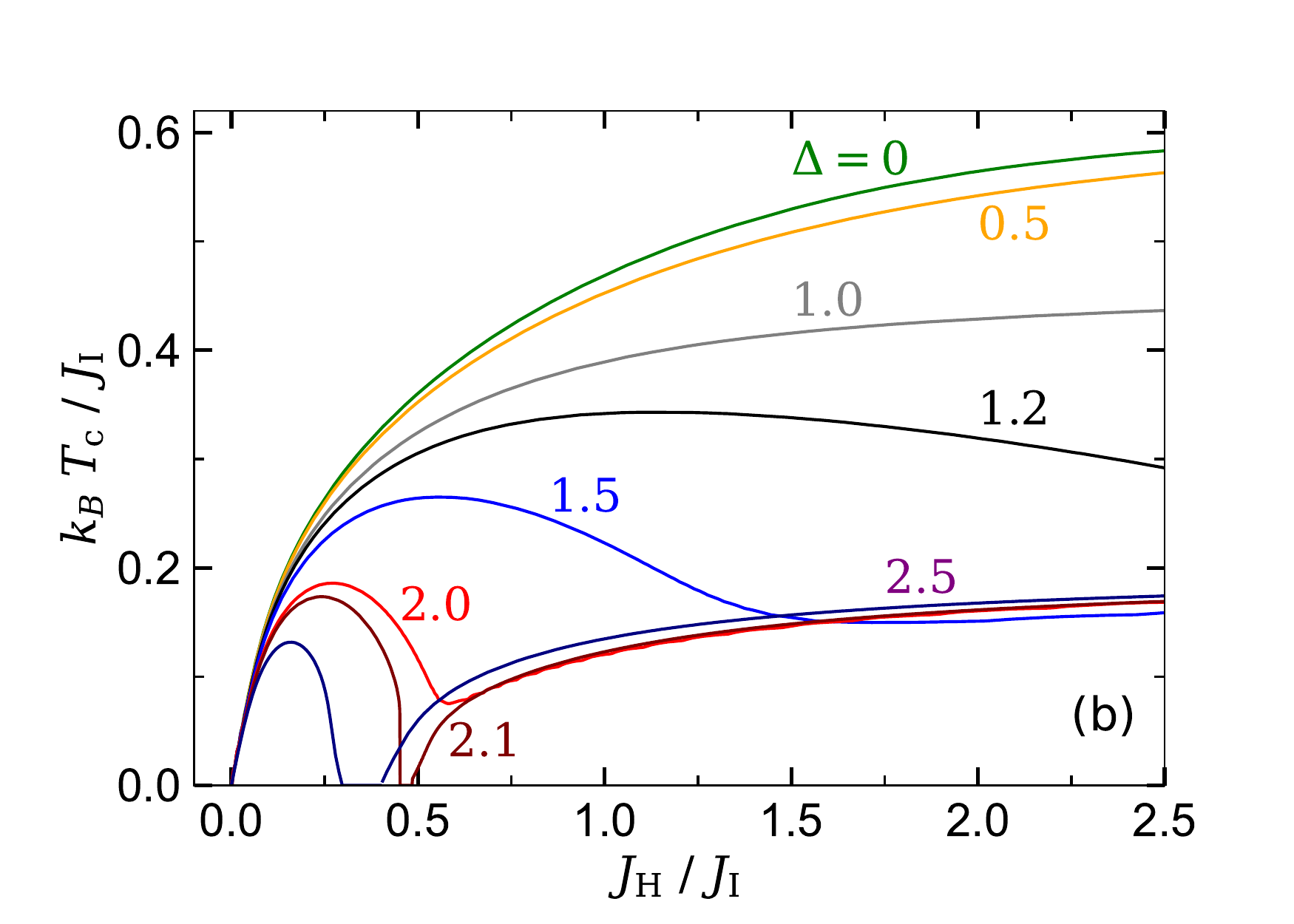}
\caption{Critical temperature of the spin-1/2 Ising-Heisenberg model as a function of the interaction  ratio $J_\text{H}/J_\text{I}$ for a few selected values of the exchange anisotropy $\Delta$ and:  (a) the martini lattice with $\gamma =1$; (b) the martini-diced lattice with $\gamma =2$.}
\label{fig:QFT_gamma12}       
\end{center}
\end{figure*}

\subsection{Critical temperature}
Bearing all this in mind, let us now discuss the dependence of the critical temperature of the spin-1/2 Ising-Heisenberg model on the 
martini and martini-diced lattice on the interaction ratio $J_\text{H}/J_\text{I}$. It is evident from Fig. \ref{fig:QFT_gamma12} that the critical temperature monotonically decreases as the interaction ratio $J_\text{H}/J_\text{I}$ is lowered until it completely vanishes at $J_\text{H}/J_\text{I}=0$ whenever the isotropic Heisenberg coupling with $\Delta=1$ or the one with the easy-axis exchange anisotropy $\Delta < 1$ is considered. On the other hand, the critical temperature displays for the Heisenberg coupling with easy-plane exchange anisotropies of moderate strength $1<\Delta<2$ an outstanding nonmonotonous dependence with a shallow minimum and a broad maximum. The critical temperature of the spin-1/2 Ising-Heisenberg model on the martini lattice strikingly touches absolute zero temperature at the triple coexistence point $[\Delta,J_\text{H}/J_\text{I}]=[2,1/2]$ of the ground states CFP, QFP and FRU in opposite to the spin-1/2 Ising-Heisenberg model on the martini-diced lattice, which has a nonzero critical temperature above the relevant triple coexistence point. The critical temperature of the spin-1/2 Ising-Heisenberg model on the martini and martini-diced lattice for the stronger easy-plane exchange anisotropies $\Delta>2$ generally displays a dome-like behavior inherent to the spontaneously ordered CFP emergent at sufficiently low values of the interaction ratio $J_\text{H}/J_\text{I}$, then the gap with zero critical temperature appears in a relative narrow range of the interaction ratio $J_\text{H}/J_\text{I}$ of moderate strength due to existence of the disordered ground state FRU and finally, the critical temperature rises steadily at higher values of the interaction ratio $J_\text{H}/J_\text{I}$ pertinent to the spontaneously ordered QFP.
Generally, the critical temperature of the spin-1/2 Ising-Heisenberg model on the martini-diced lattice is shifted towards higher values in comparison with that one of the spin-1/2 Ising-Heisenberg model on the martini lattice, because the martini-diced lattice has higher connectivity of sites with respect to the martini lattice reflected in a two-times stronger effective coupling.

\begin{figure}
\begin{center}
\includegraphics[scale=0.45,trim=10 19 20 0, clip]{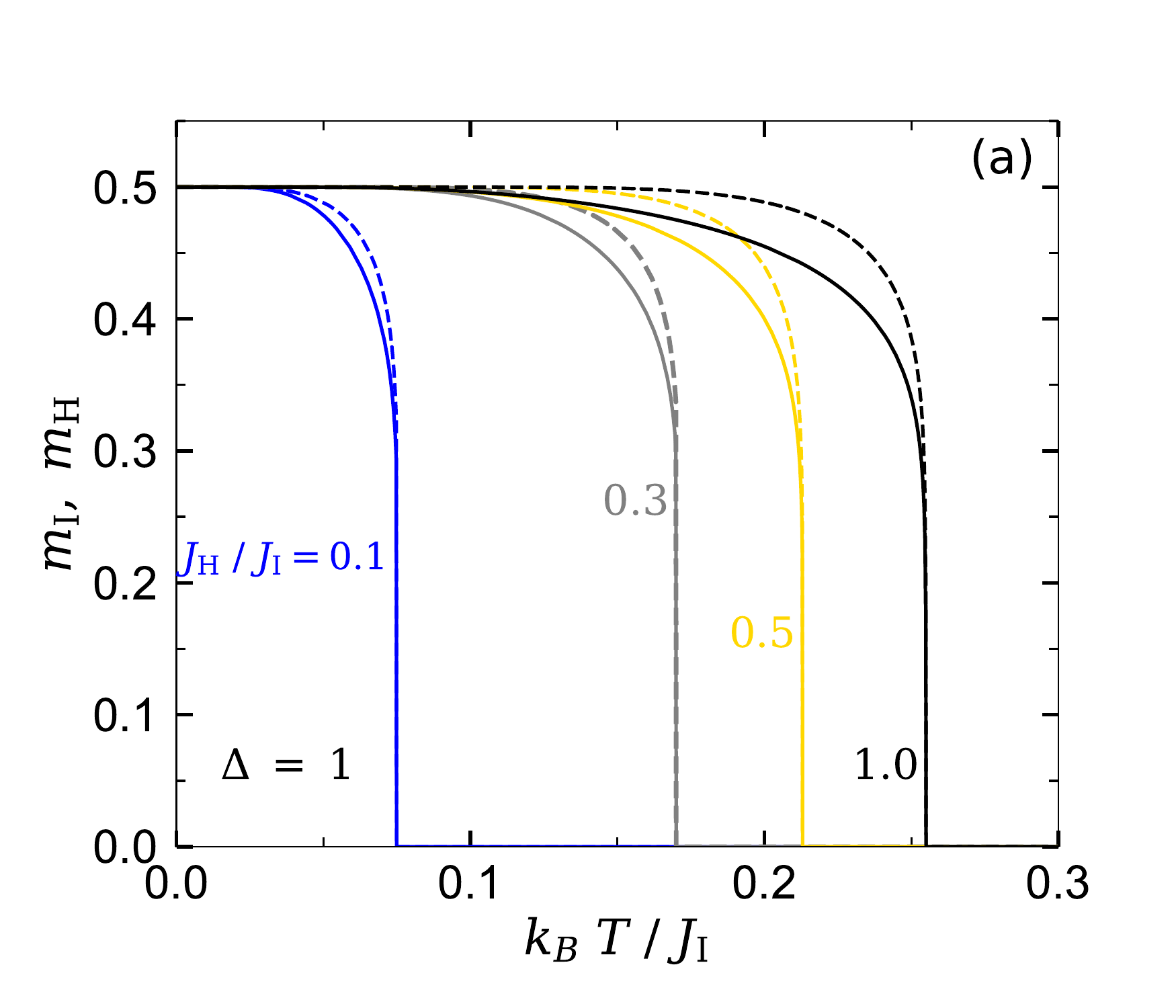}
\includegraphics[scale=0.45,trim=10 19 20 20, clip]{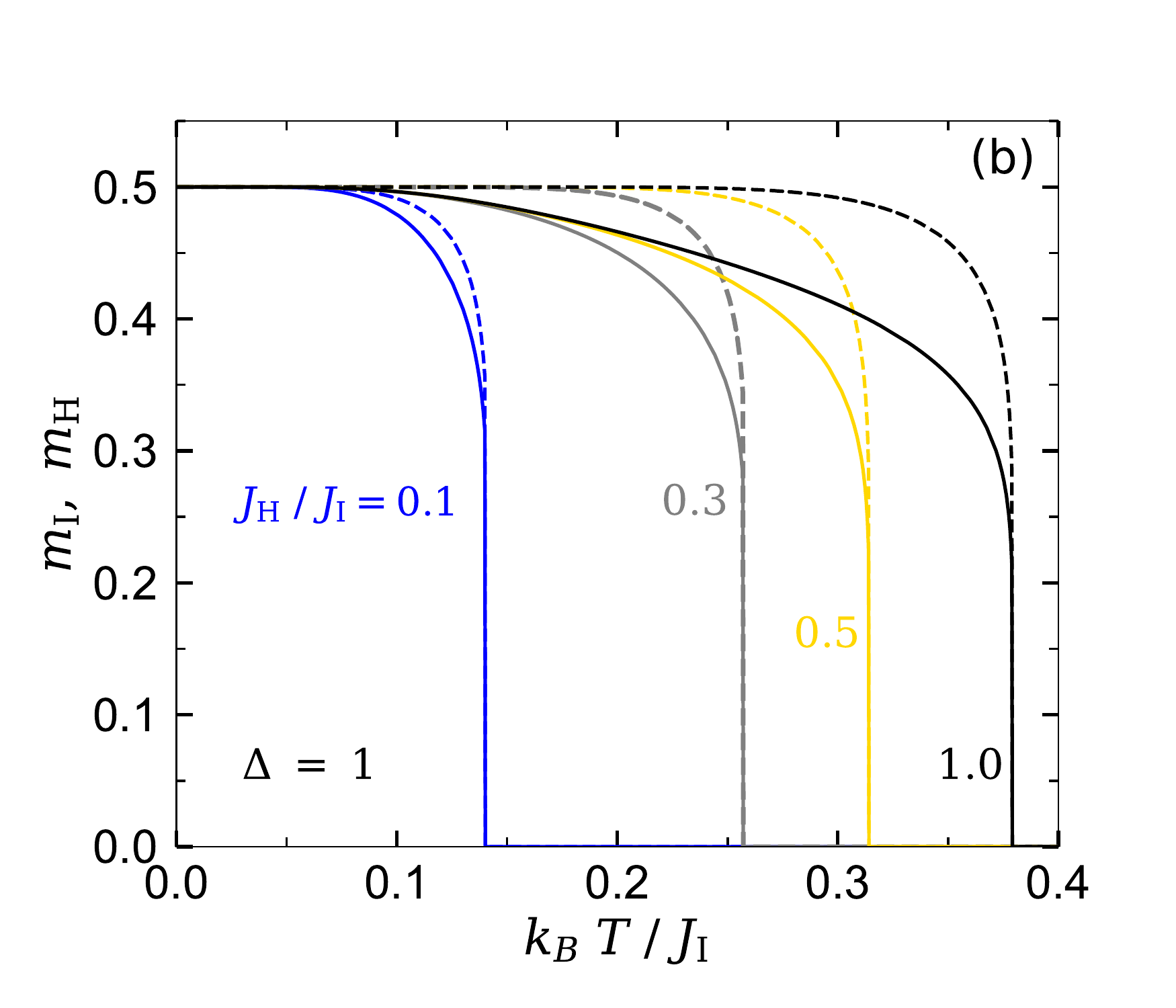}
\caption{Temperature dependencies of the spontaneous magnetization $m_\text{I}$ of the Ising spins (broken lines) and the spontaneous magnetization $m_\text{H}$ of the Heisenberg spins (solid lines) for the fixed value of $\Delta=1$ and several values of the interaction ratio $J_\text{H}/J_\text{I}$ by considering: (a) the martini lattice with $\gamma =1$, (b) the martini-diced lattice with $\gamma =2$.}
\label{fig:Mag_Delta1}       
\end{center}
\end{figure}

\subsection{Magnetization}
 Temperature dependencies of the spontaneous magnetization of the spin-1/2 Ising-Heisenberg model on the martini and martini-diced lattices are illustrated in Fig. \ref{fig:Mag_Delta1} for several  values of the coupling ratio  $J_\text{H}/J_\text{I}$ and the particular choice of the parameter $\Delta=1$. It is quite clear from Fig. \ref{fig:Mag_Delta1}(a) that the spontaneous magnetizations of the Heisenberg and Ising spins for the spin-1/2 Ising-Heisenberg model on the martini lattice start from their saturation value in accordance with the classical ground state CFP. Moreover, the spontaneous magnetization of the Heisenberg spins generally exhibits a greater temperature-induced decline than the spontaneous magnetization of the Ising spins even though both sublattice magnetizations tend towards zero nearby a critical temperature with the same critical exponent from the standard universality class of two-dimensional Ising model. A similar behavior can be observed in the temperature dependencies of the spontaneous magnetizations of the spin-1/2 Ising-Heisenberg model on the martini-diced lattice [see Fig. \ref{fig:Mag_Delta1}(b)], but the critical temperature at which both sublattice magnetizations suddenly vanish moves in agreement with previous argumentation towards higher temperatures.

\begin{figure}
\begin{center}
\includegraphics[scale=0.45,trim=10 19 20 0, clip]{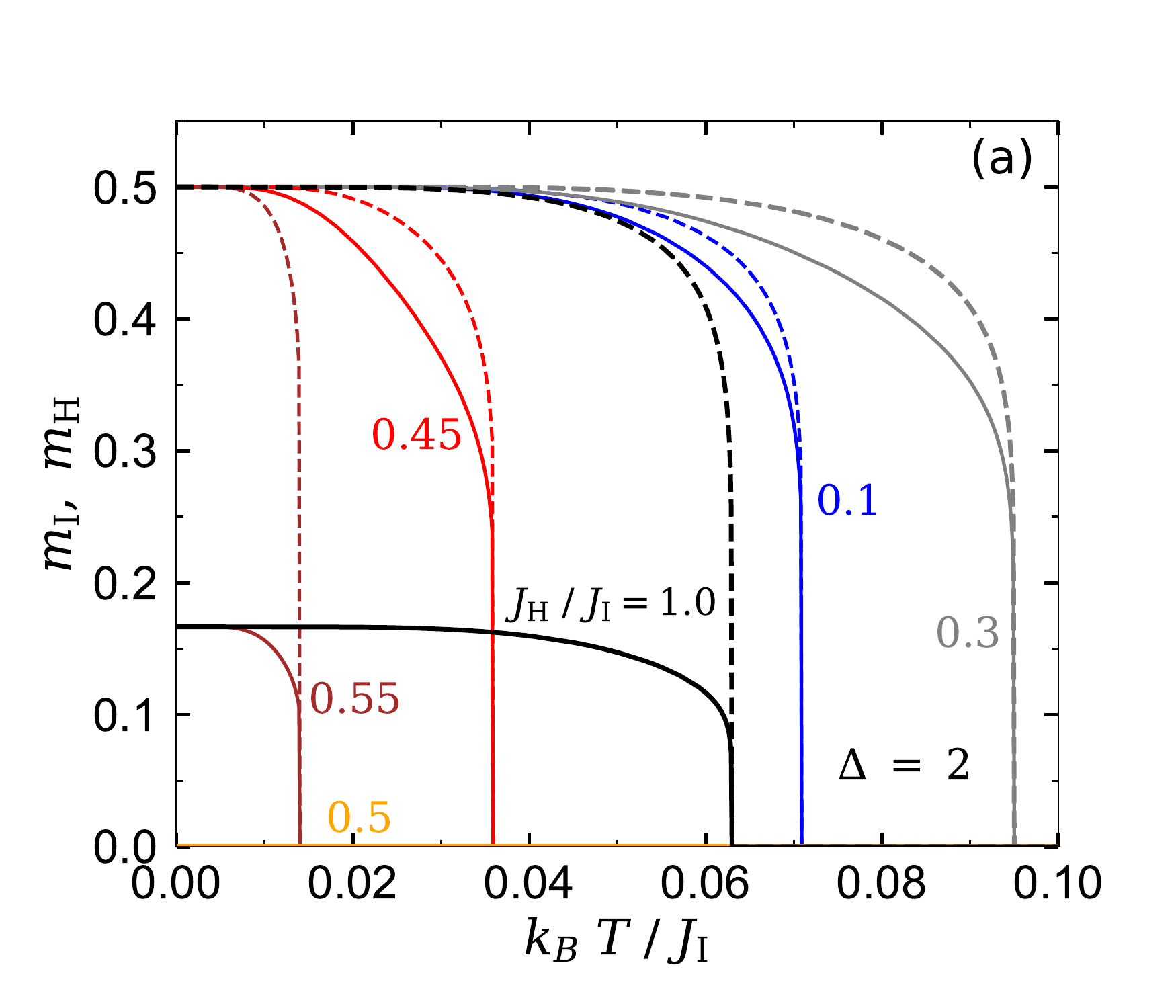}
\includegraphics[scale=0.45,trim=10 19 20 20, clip]{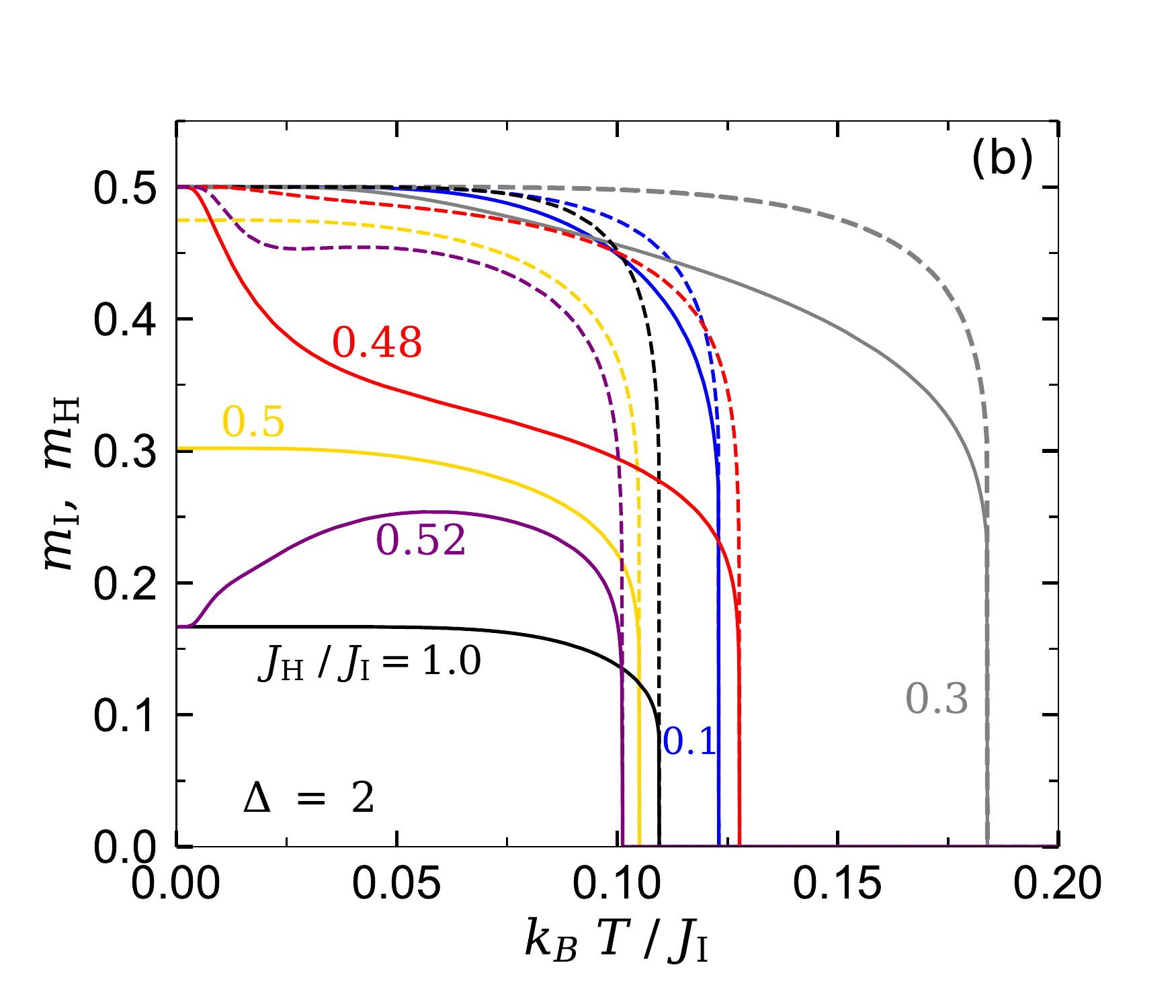}
\caption{Temperature dependencies of the spontaneous magnetization $m_\text{I}$ of the Ising spins (broken lines) and the spontaneous magnetization $m_\text{H}$ of the Heisenberg spins (solid lines) for the fixed value of $\Delta=2$ and several values of the interaction ratio $J_\text{H}/J_\text{I}$ by considering: (a) the martini lattice with $\gamma =1$, (b) the martini-diced lattice with $\gamma =2$.}
\label{fig:Mag_Delta2}       
\end{center}
\end{figure}

Next, temperature dependencies of the spontaneous magnetization of the spin-1/2 Ising-Heisenberg model on the martini and martini-diced lattices are depicted in Fig. \ref{fig:Mag_Delta2} for a few selected values of the interaction ratio  $J_\text{H}/J_\text{I}$ and the easy-plane exchange anisotropy $\Delta=2$. A closer inspection of Fig. \ref{fig:Mag_Delta2}(a) for the spin-1/2 Ising-Heisenberg model on the martini lattice reveals that the spontaneous magnetization of the Heisenberg spins starts for sufficiently large values of the interaction ratio $J_\text{H}/J_\text{I}>0.5$ from the unsaturated value $m_\text{H}=1/6$, which bears evidence of the ground state QFP. Furthermore, both spontaneous magnetizations are equal zero at the triple coexistence point $[\Delta,J_\text{H}/J_\text{I}]=[2, 1/2]$ for any temperature. It is quite obvious from Fig. \ref{fig:Mag_Delta2}(b) that the spontaneous magnetizations of the Heisenberg  and Ising spins for the spin-1/2 Ising-Heisenberg model on the martini-diced lattice exhibits strikingly distinct behavior with respect to the previous case. First of all, the spontaneous magnetization of the Heisenberg spins $m_\text{H}$ displays sharply descending S-shaped temperature dependence for the interaction ratio $J_\text{H}/J_\text{I}\lesssim 0.5$, while the spontaneous magnetization of the Ising spins $m_\text{I}$ initially gradually decreases and suddenly vanishes just nearby the critical temperature. Even more intriguing behavior of the spontaneous magnetizations can be observed for the interaction ratio $J_\text{H}/J_\text{I}\gtrsim 0.5$. Under this condition, the spontaneous magnetization of the Heisenberg spins $m_\text{H}$ exhibits a marked temperature-induced rise from the initial value $m_\text{H}=1/6$ until it reaches a relatively wide maximum and finally it suddenly vanishes at a critical temperature. On the other hand, the spontaneous magnetization of the Ising spins $m_\text{I}$ shows a remarkable descending S-shaped temperature dependence starting from its saturation value $m_\text{I}=1/2$. If the coupling ratio is much higher than $J_\text{H}/J_\text{I} \gg 0.5$, then, the spontaneous magnetization of the Heisenberg spins $m_\text{H}$ is kept constant $m_\text{H}=1/6$ over a relatively wide temperature interval including low up to moderate temperatures and it suddenly drops to zero just in a close vicinity of the critical temperature. 

\begin{figure*}
\begin{center}
\includegraphics[scale=0.55,trim=0 0 10 10, clip]{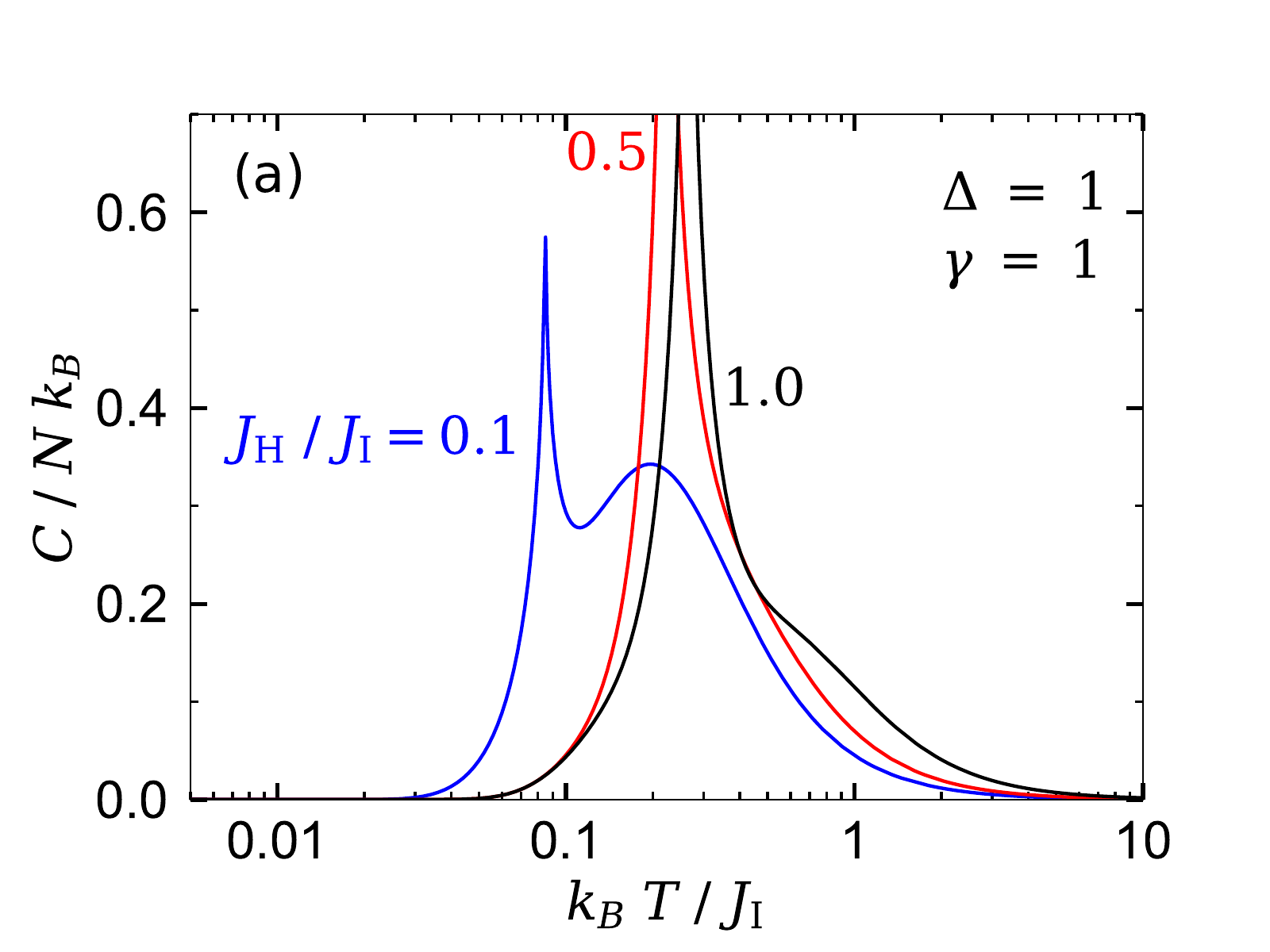}
\includegraphics[scale=0.55,trim=0 0 10 10, clip]{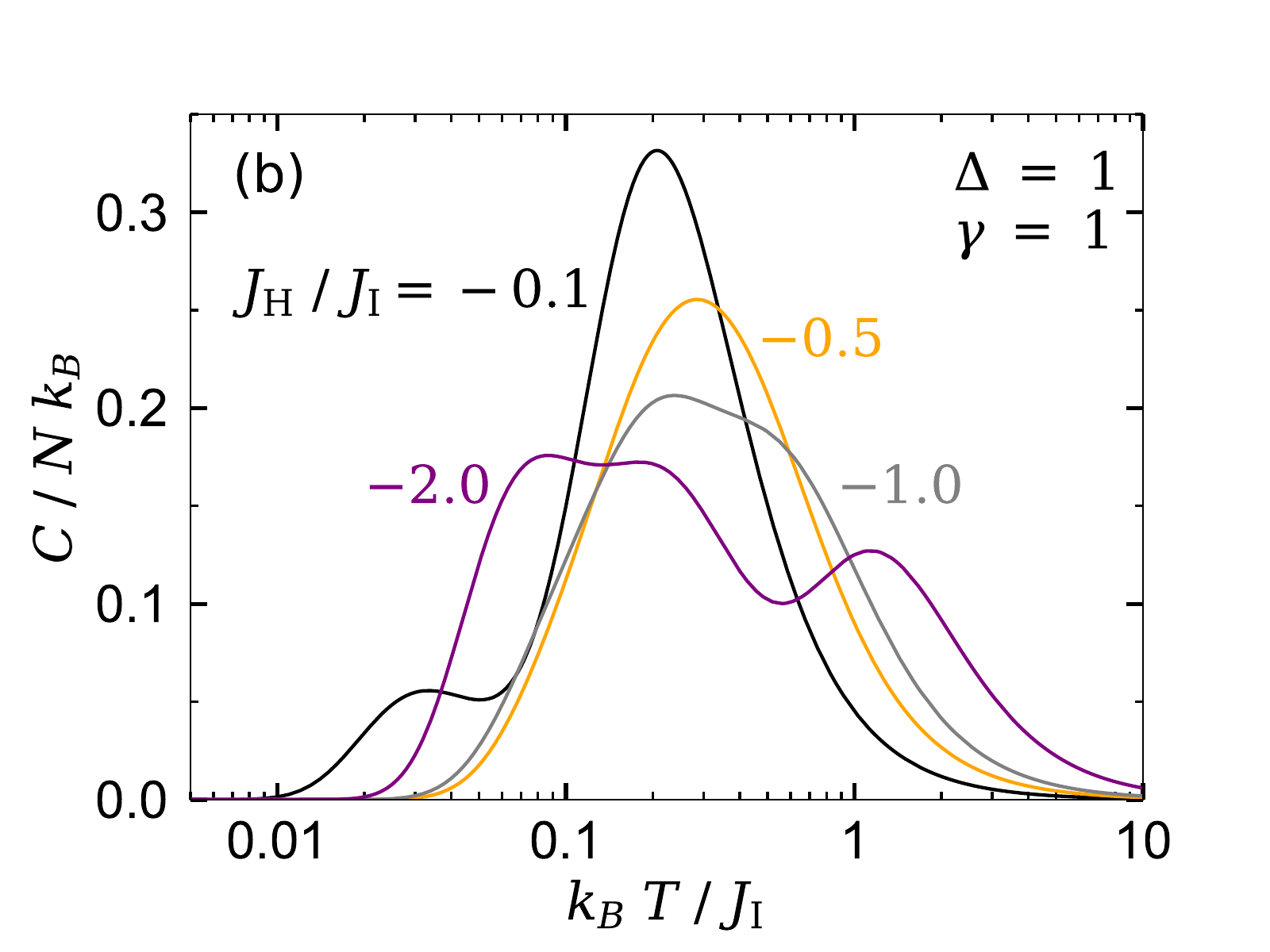}
\includegraphics[scale=0.55,trim=0 0 10 10, clip]{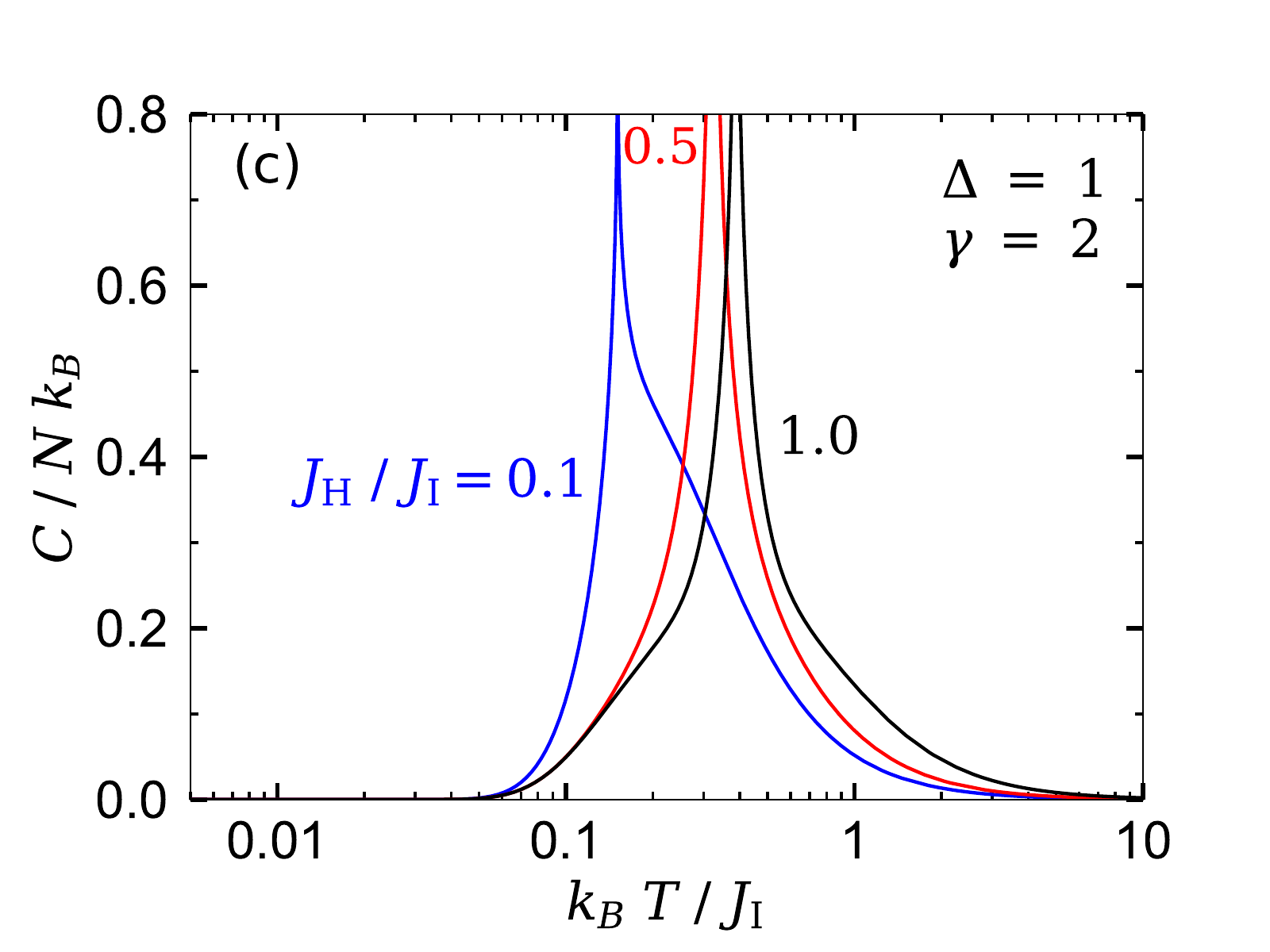}
\includegraphics[scale=0.55,trim=0 0 10 10, clip]{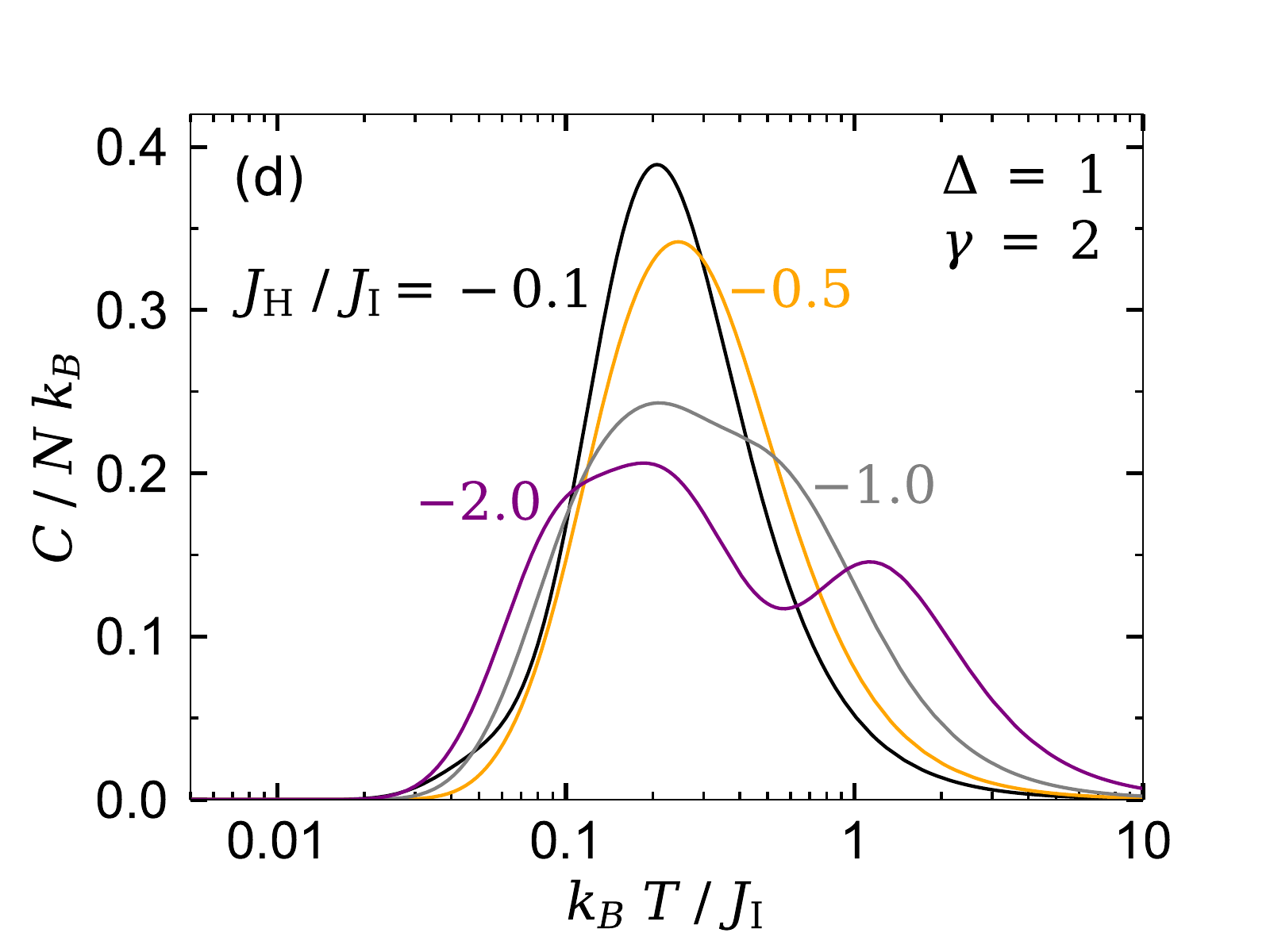}
\caption{Temperature dependencies of the specific heat of the spin-1/2 Ising-Heisenberg model for a few selected values of the interaction ratio  $J_\text{H}/J_\text{I}$ and the fixed value of the parameter $\Delta=1$ (the isotropic case) by considering the specific case defined on: (a)-(b) the martini lattice with $\gamma =1$, (c)-(d) the martini-diced lattice with $\gamma =2$.}
\label{fig:SHeat_gamma1}       
\end{center}
\end{figure*}
 
\subsection{Specific heat}
Finally, let us turn our attention to a detailed analysis of the temperature dependencies of the specific heat of the spin-1/2 Ising-Heisenberg model on two considered martini-type lattices. The zero-field specific heat is plotted in Fig. \ref{fig:SHeat_gamma1} against temperature in a semilogarithmic scale by assuming the fixed value of the parameter $\Delta=1$ and a few selected values of the interaction ratio $J_\text{H}/J_\text{I}$ for the martini lattice [Figs. \ref{fig:SHeat_gamma1}(a) and \ref{fig:SHeat_gamma1}(b)] as well as the martini-diced lattice [Figs. \ref{fig:SHeat_gamma1}(c) and \ref{fig:SHeat_gamma1}(d)]. It follows from Fig. \ref{fig:SHeat_gamma1}(a) and (c) that the specific heat exhibits a single logarithmic divergence from the standard Ising universality class when considering the ferromagnetic Heisenberg interaction $J_\text{H}/J_\text{I}>0$, which relates to a continuous phase transition from the spontaneously ordered CFP to the disordered paramagnetic phase. On the contrary, the logarithmic divergence is evidently absent in thermal variations of the specific heat depicted in Fig. \ref{fig:SHeat_gamma1}(b) and (d) for the antiferromagnetic Heisenberg interaction $J_\text{H}/J_\text{I}<0$, which convincingly evidences a lack of any spontaneous long-range order at low enough temperatures and is consistent with presence of the disordered ground state FRU. In consequence of that, one detects in this frustrated parameter region temperature dependencies of the specific heat with either single, double or triple round maximum, which may be under certain circumstances superimposed on each other and thus responsible for irregular shape of the round maximum.  

\begin{figure*}
\begin{center}
\includegraphics[scale=0.55,trim=0 0 10 10, clip]{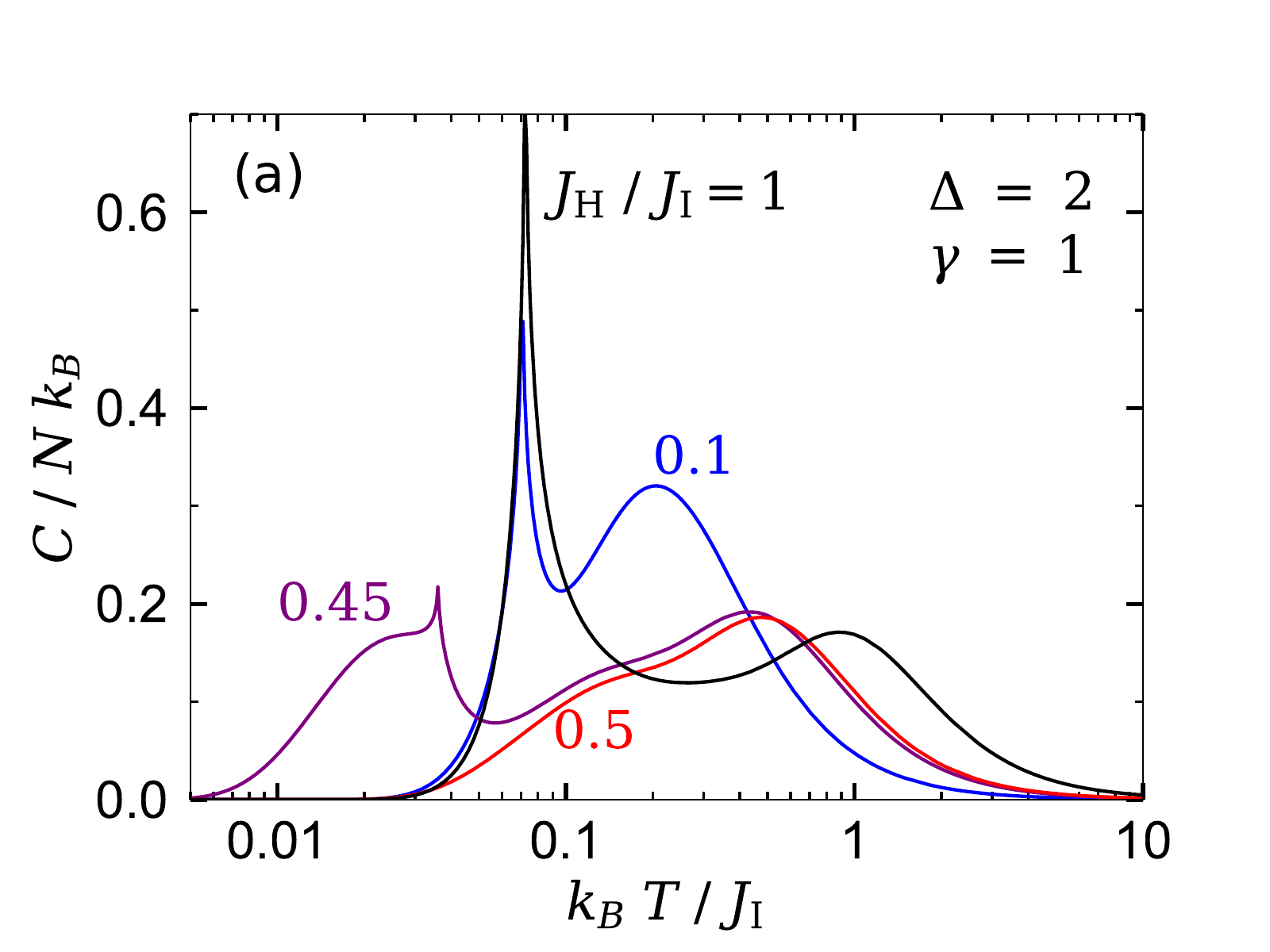}
\includegraphics[scale=0.55,trim=0 0 10 10, clip]{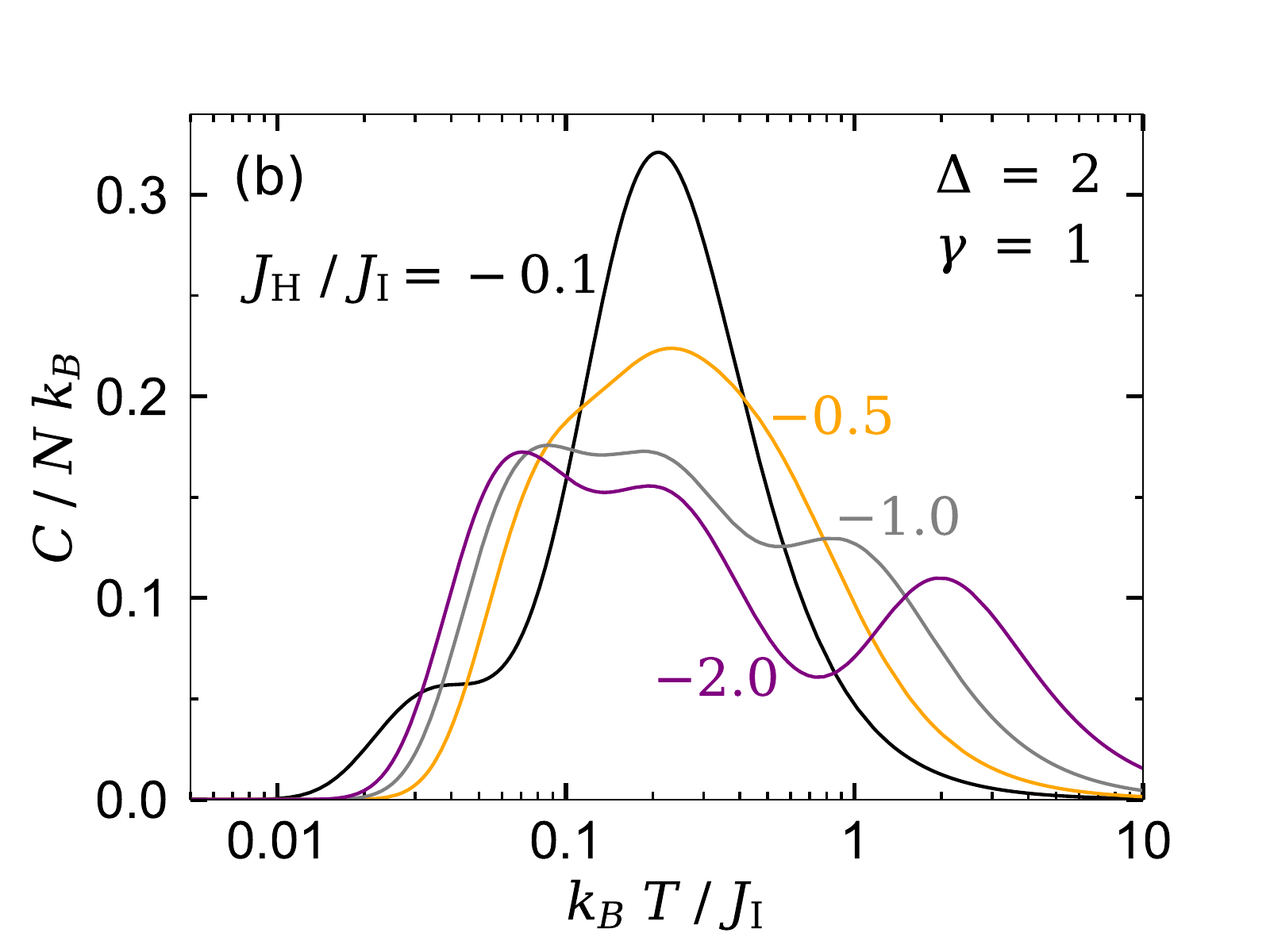}
\includegraphics[scale=0.55,trim=0 0 10 10, clip]{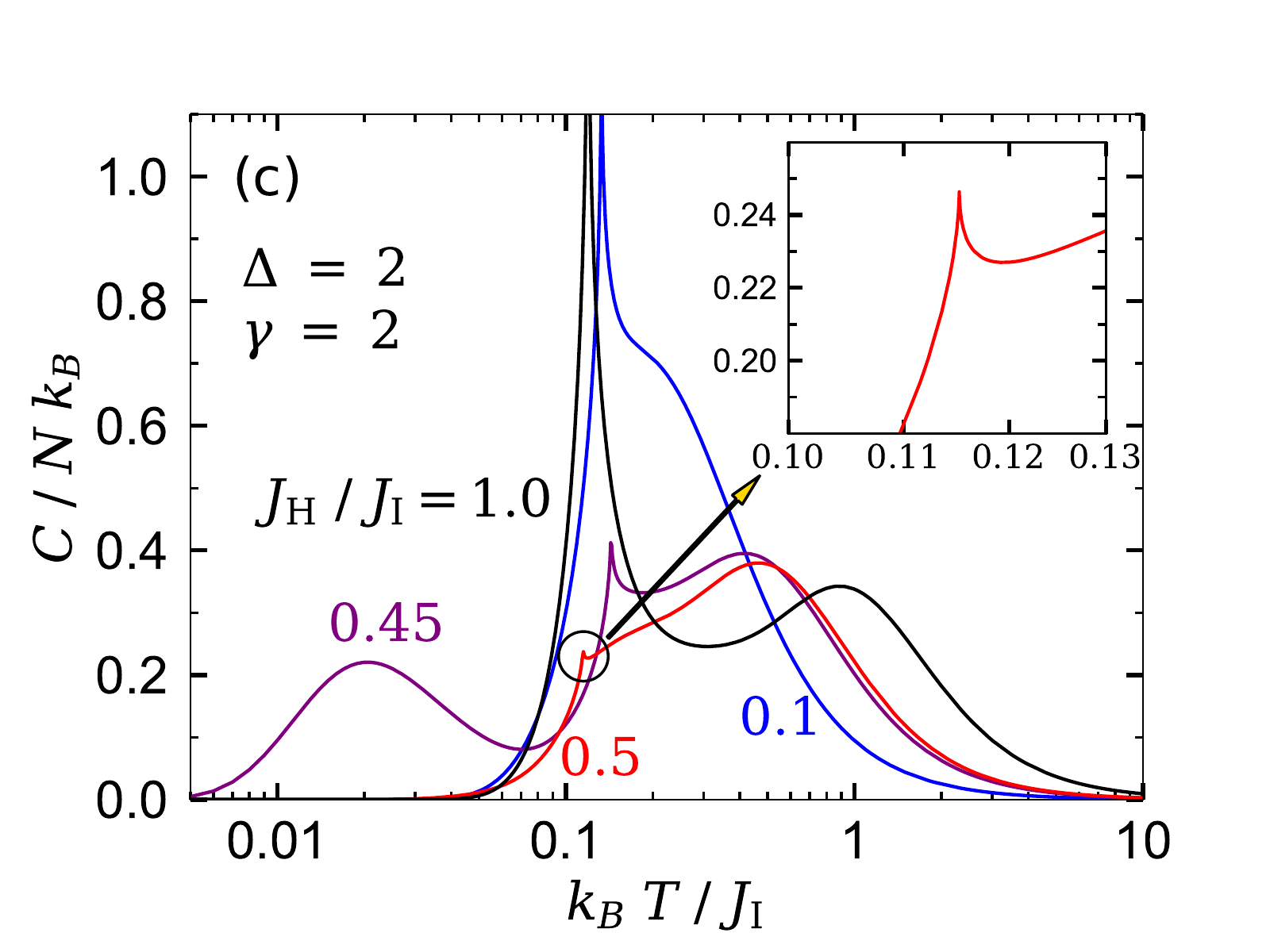}
\includegraphics[scale=0.55,trim=0 0 10 10, clip]{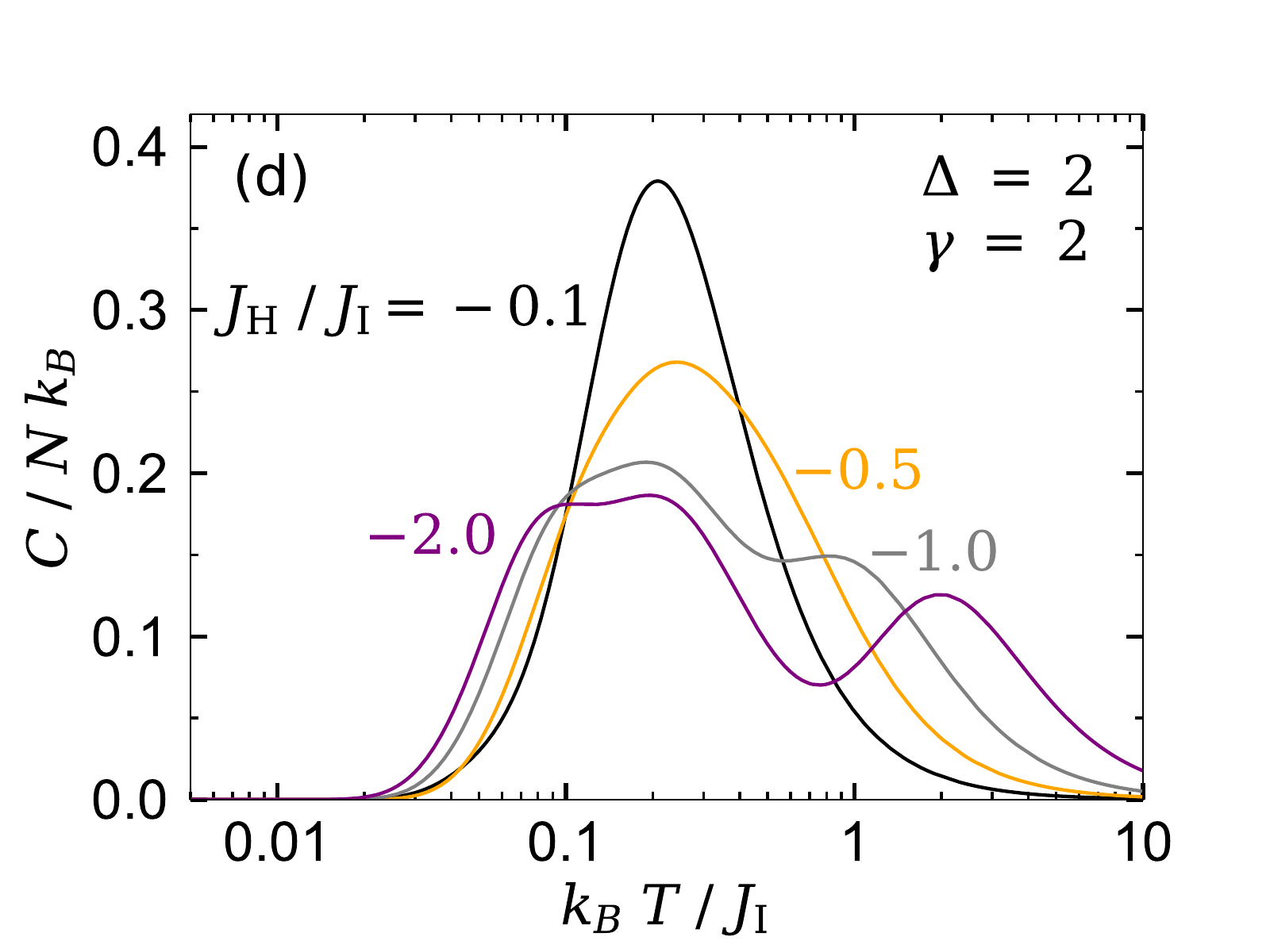}
\caption{Temperature dependencies of the specific heat of the spin-1/2 Ising-Heisenberg model for a few selected values of the interaction ratio  $J_\text{H}/J_\text{I}$ and the fixed value of the parameter $\Delta=2$ by considering the specific case defined on: (a)-(b) the martini lattice with $\gamma =1$, (c)-(d) the martini-diced lattice with $\gamma =2$.}
\label{fig:SHeat_gamma2}       
\end{center}
\end{figure*}

Furthermore, temperature dependencies of the specific heat of the spin-1/2 Ising-Heisenberg model on the martini and martini-diced lattices are displayed in Fig. \ref{fig:SHeat_gamma2} for a few selected values of the interaction ratio $J_\text{H}/J_\text{I}$ and the specific choice of the easy-plane exchange anisotropy $\Delta=2$. As one can see, the specific heat exhibits for this highly anisotropic case qualitatively similar thermal behavior as the formerly discussed isotropic case $\Delta=1$ if the ferromagnetic Heisenberg interaction is sufficiently weak $J_\text{H}/J_\text{I} \gtrsim 0$, c.f. blue curves in Figs. \ref{fig:SHeat_gamma1}(a),(c) and \ref{fig:SHeat_gamma2}(a),(c). This coincidence is attributable to the same character of the spontaneously ordered ground state CFP and its elementary excitations being the same for both considered cases at low enough temperatures. However, temperature variations of the specific heat do not coincide at higher values of the ferromagnetic Heisenberg interaction $J_\text{H}/J_\text{I} \gtrsim 0.5$, because characteristic features of the specific heat for the exchange anisotropy $\Delta=2$ are fundamentally influenced by a zero-temperature phase transition between CFP and QFP that is fully absent in the former isotropic case $\Delta=1$. Owing to this fact, the round low-temperature maximum on an ascending tail of the logarithmic divergence of the specific heat develops for the particular case $J_\text{H}/J_\text{I} = 0.45$, which can be explained in terms of low-energy excitation from CFP towards QFP.  Another interesting observation is that the logarithmic singularity completely disappears when the spin-1/2 Ising-Heisenberg model on the martini lattice is driven to the triple point $[\Delta,J_\text{H}/J_\text{I}]=[2, \frac{1}{2}]$ [see the red line in Fig. \ref{fig:SHeat_gamma2}(a)], while it is still preserved for the spin-1/2 Ising-Heisenberg model on the martini-diced lattice [see the red line in Fig. \ref{fig:SHeat_gamma2}(c) and its inset]. It is also quite evident from Fig. \ref{fig:SHeat_gamma2}(a) and (c) that the critical amplitude of the logarithmic divergence gradually decreases as one reaches the ground-state phase boundary between CFP and QFP. As a matter of fact, the pronounced logarithmic divergence of the specific heat can be repeatedly observed in Fig. \ref{fig:SHeat_gamma2}(a) and (c) for the interaction ratio 
$J_\text{H}/J_\text{I} = 1$, which drives the spin-1/2 Ising-Heisenberg model on the martini and martini-diced lattices towards the other spontaneously ordered ground state QFP [see black lines in Fig. \ref{fig:SHeat_gamma2}(a) and (c)]. As far as the particular case with the antiferromagnetic Heisenberg coupling $J_\text{H}/J_\text{I}<0$ is concerned, the specific heat exhibits a smooth temperature dependence with one up to three more or less discernible round maxima [see Fig. \ref{fig:SHeat_gamma2}(b) and (d)]. A smooth thermal dependence without any obvious singularity is consistent with a macroscopic degeneracy of the ground state FRU whose disordered nature leads to absence of the spontaneous long-range order. It can be concluded that the specific heat shows a logarithmic singularity just if the ground state can be ascribed to a spontaneously long-range ordered CFP or QFP.
\begin{figure*}
\begin{center}
\includegraphics[scale=0.55,trim=5 5 10 10, clip]{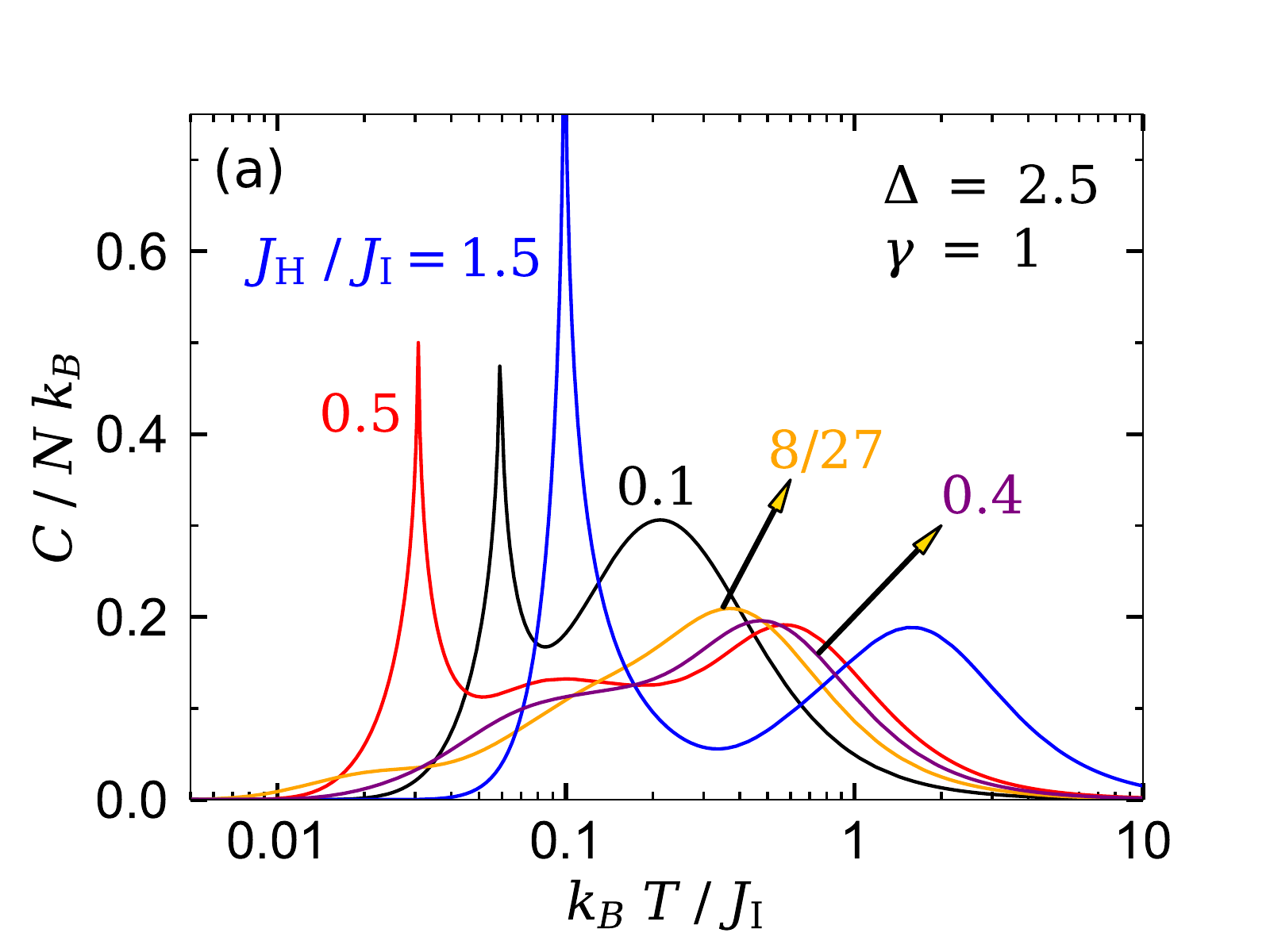}
\includegraphics[scale=0.55,trim=10 5 10 10, clip]{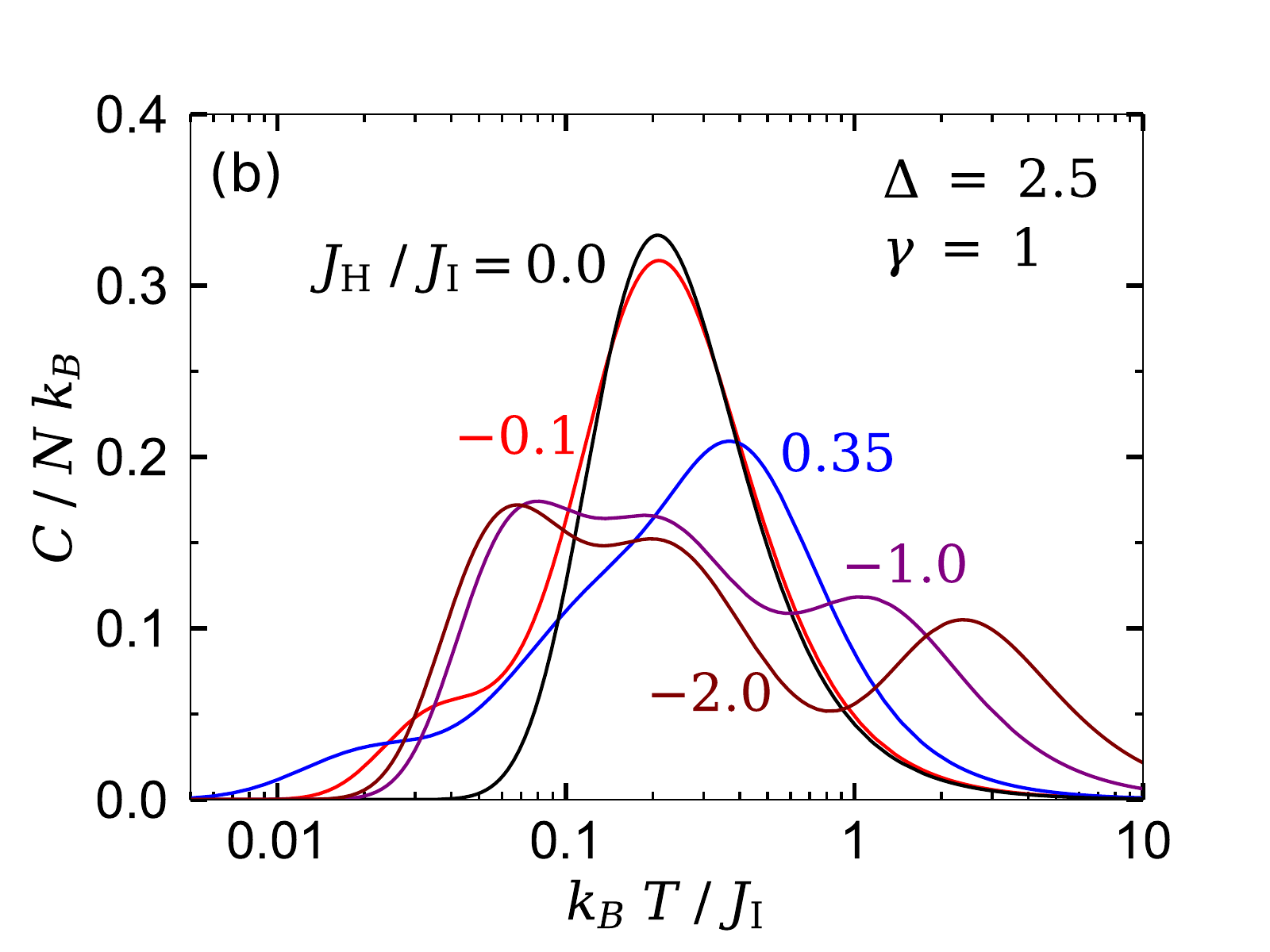}
\includegraphics[scale=0.55,trim=5 5 10 10, clip]{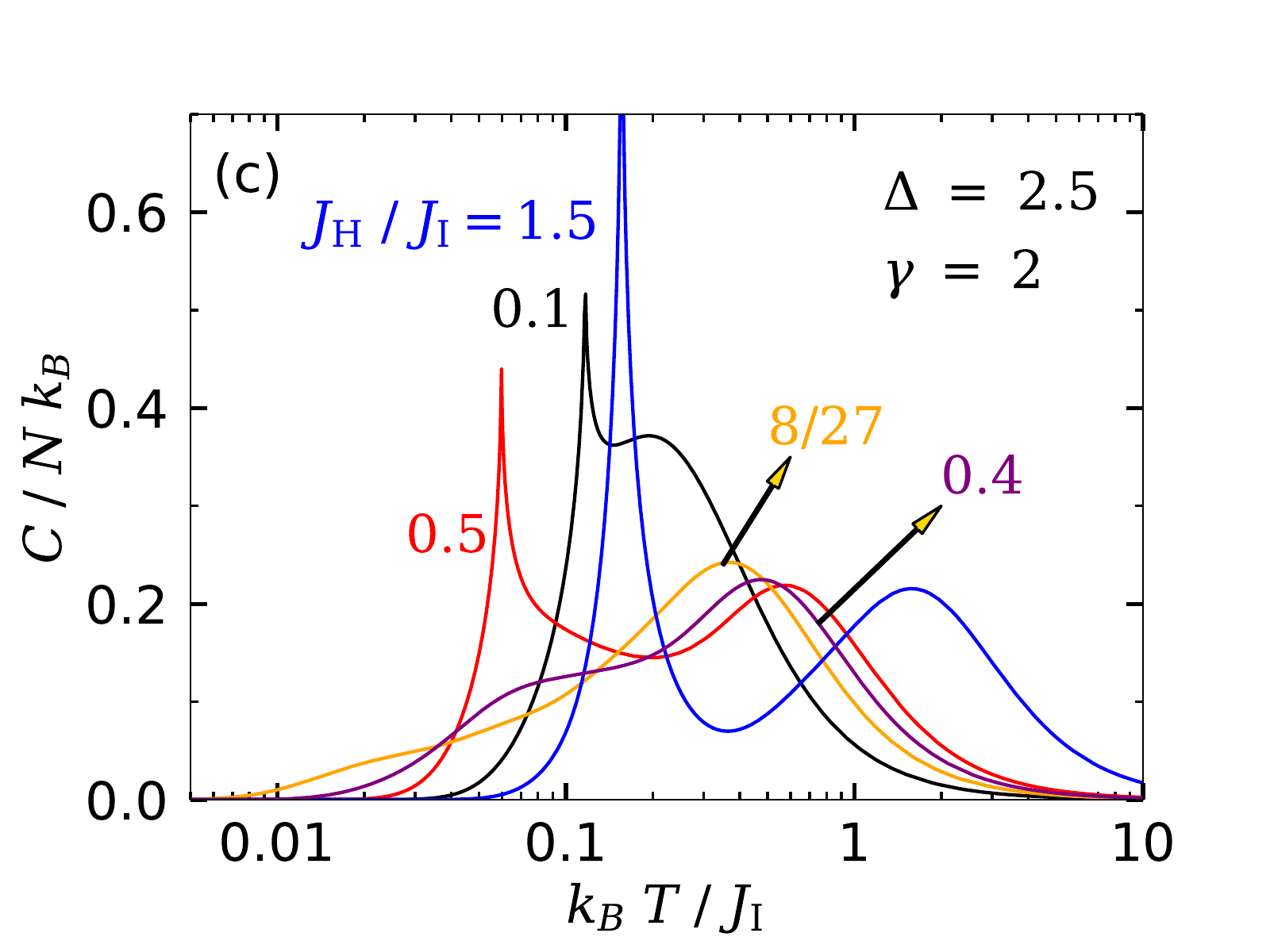}
\includegraphics[scale=0.55,trim=10 5 10 10, clip]{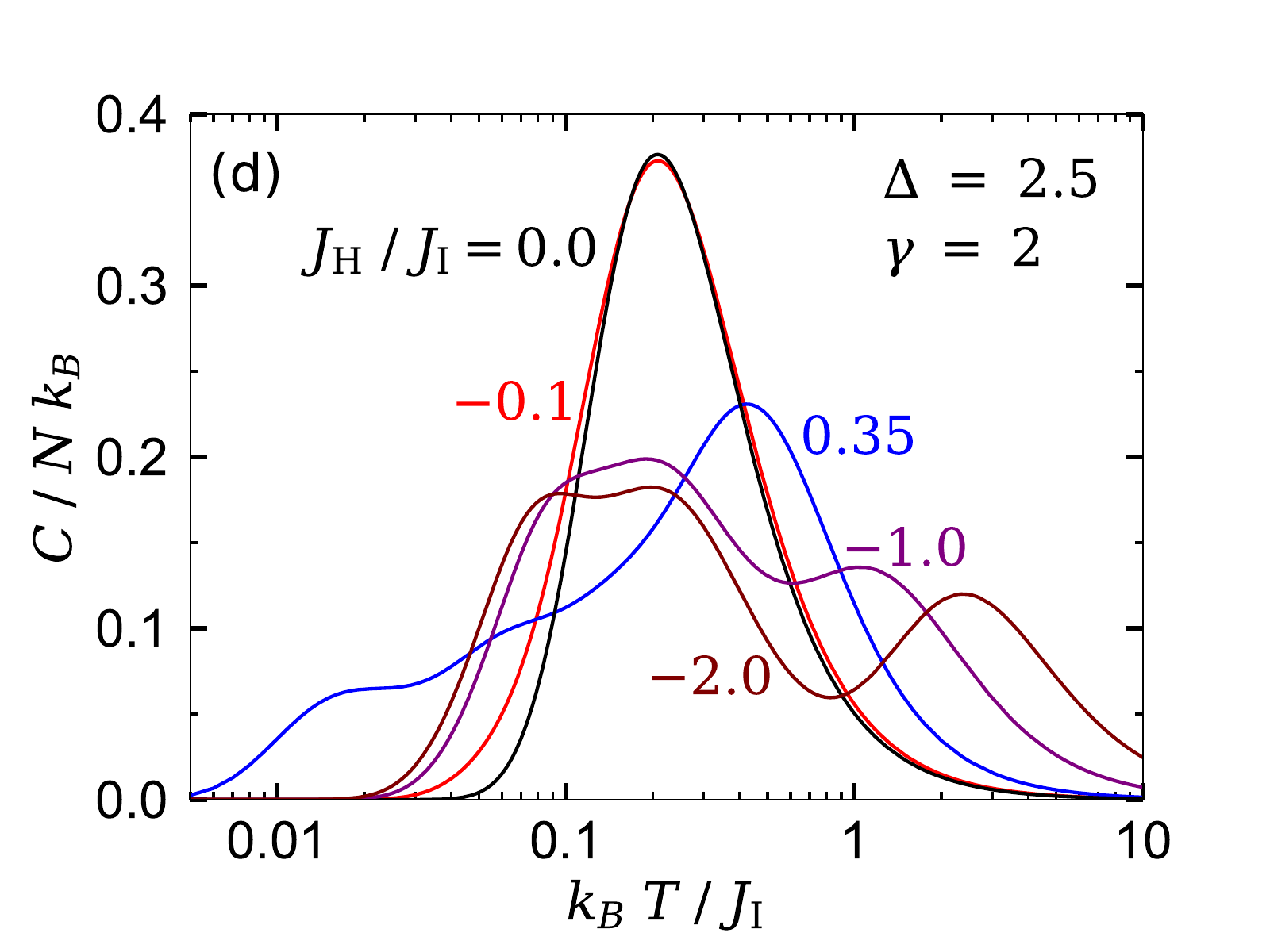}
\caption{Temperature dependencies of the specific heat of the spin-1/2 Ising-Heisenberg model for a few selected values of the interaction ratio  $J_\text{H}/J_\text{I}$ and the fixed value of the parameter $\Delta = 2.5$ by considering the specific case defined on: (a)-(b) the martini lattice with $\gamma = 1$, (c)-(d) the martini-diced lattice with $\gamma = 2$.}
\label{fig:SHeat2c5}       
\end{center}
\end{figure*}

Last but not least, we will conclude our analysis of temperature variations of the specific heat by considering the special case of the spin-1/2 Ising-Heisenberg model on the martini and martini-diced lattices with the fixed value of the exchange anisotropy $\Delta=2.5$. Under this condition, a logarithmic divergence observable in thermal dependencies of the specific heat corresponds either to a breakdown of the spontaneously ordered QFP (for $J_\text{H}/J_\text{I}>2/5$) or CFP (for $0<J_\text{H}/J_\text{I}<8/27$), whereas the specific heat is free of any divergence for $J_\text{H}<0$ due to the disordered ground state FRU (see Fig. \ref{fig:SHeat2c5}). All the aforementioned thermal dependencies have quite similar features as described previously and our attention will be henceforth restricted to a narrow range of the interaction ratio $J_\text{H}/J_\text{I} \in (8/27,2/5)$, which promotes presence of the disordered ground state FRU in spite of the purely ferromagnetic character of both interaction constants. It is clear from Fig. \ref{fig:SHeat2c5}(a) and (c) that the specific heat shows a low-temperature hump superimposed on a round maximum for two boundary values $J_\text{H}/J_\text{I} = 8/27$ and $2/5$, which correspond to the ground-state phase boundaries CFP-FRU and QFP-FRU, respectively. As far as a typical behavior of the disordered ground state FRU in the ferromagnetic parameter space is concerned, one detects an outstanding corrugated profile in a temperature dependence of the specific heat due to superimposition of two humps on a round maximum [see the curves for $J_\text{H}/J_\text{I} = 0.35$ in Fig. \ref{fig:SHeat2c5}(b) and (d)].

\section{Conclusions}\label{conclusions}

In the present paper, we have exactly solved the spin-1/2 Ising-Heisenberg model on two related martini-type lattices through the generalized star-triangle transformation, which establishes a rigorous mapping correspondence to an effective spin-1/2 Ising model on a triangular lattice. The spin-1/2 Ising-Heisenberg model on the martini and martini-diced lattices accordingly shows presence or absence of the spontaneous long-range order at low enough temperatures depending on whether the effective interaction is ferromagnetic or antiferromagnetic. In particular, we have investigated with the help of this exact mapping equivalence the ground-state and finite-temperature phase diagrams, as well as, temperature dependencies of the spontaneous magnetization and specific heat. 

It has been evidenced that the spin-1/2 Ising-Heisenberg model on two considered martini-type lattices exhibits two spontaneously long-range ordered ground states (CFP and QFP) and one disordered ground state (FRU) with a nonzero residual entropy. A presence of the spontaneous ordering within CFP and QFP has been verified through a detailed examination of temperature dependencies of the sublattice magnetization of the Ising and Heisenberg spins, the latter of which enables to discern the classical ferromagnetic ordering CFP from the quantum one QFP  being subject to a quantum reduction of the magnetization due to a symmetric quantum superposition of three uud states of the Heisenberg trimers. Although existence of the disordered ground state FRU for the antiferromagnetic Heisenberg interaction $J_\text{H}<0$ introducing a geometric spin frustration is quite reasonable and somewhat expected, we have surprisingly found out that the same disordered ground state FRU may be curiously stabilized also by an unconventional spin frustration arising from a competition of the purely ferromagnetic Ising and Heisenberg interactions of easy-axis and easy-plane type, respectively. 

An emergence of the unconventional disordered phase FRU in the spin-1/2 Ising-Heisenberg model on martini and martini-diced lattices with pure ferromagnetic interactions represents the most outstanding finding of the present work, which has not been reported in the literature yet. It should be emphasized that a mutual competition between the easy-axis and easy-plane ferromagnetic interactions on some 2D Ising-Heisenberg close-packed lattice represents necessary but not sufficient condition for presence of the disordered phase FRU emergent due to the peculiar spin frustration. For a comparison, the spin-1/2 Ising-Heisenberg model on a triangulated kagom\'e lattice \cite{str08,yao08} or two related triangles-in-triangles lattices \cite{cis13} do not exhibit this peculiar feature. Although the same finding would also emerge for the spin-1/2 Ising-Heisenberg model on an expanded kagom\'e lattice \cite{roj19}, the respective parameter space has not been comprehensively explored so far.

Ultimately, it has been shown that the specific heat of the spin-1/2 Ising-Heisenberg model on the martini and martini-diced lattice with the isotropic ferromagnetic Heisenberg interaction ($J_\text{H}>0, \Delta=1$) always displays a pronounced logarithmic divergence from the standard Ising universality class due to the classical spontaneously long-range ordering CFP. In contrast to this the same model with the isotropic antiferromagnetic Heisenberg interaction ($J_\text{H}<0, \Delta=1$) always shows a smooth temperature dependence of the specific heat with one up to three round maxima and without any singularity due to a disordered character of the macroscopicallly degenerate ground state FRU. Although the same conclusion remains valid also for the spin-1/2 Ising-Heisenberg model on the martini-type lattices with the antiferromagnetic Heisenberg interaction and the exchange anisotropy $\Delta \neq 1$, the particular case with the ferromagnetic Heisenberg interaction of easy-plane type ($J_\text{H}>0, \Delta>1$) displays a more diverse thermal behavior of the specific heat. Under this condition, the logarithmic divergence of the specific heat can be found whenever the spin-1/2 Ising-Heisenberg model on the martini-type lattices is driven either to the CFP or QFP, while the round maximum is superimposed on an ascending tail of the specific-heat divergence if the model is driven sufficiently close to the ground-state phase boundary between CFP and QFP. The most spectacular temperature dependencies of the specific heat can be found in a vicinity of the triple coexistence point of the three ground states CFP, QFP and FRU, whereby the spin-1/2 Ising-Heisenberg model on the martini-diced lattice displays a peculiar spontaneous order at the triple point in opposite to the analogous model defined on a simple martini lattice. 

\section*{Acknowledgments}

Hamid Arian Zad acknowledges the receipt of the grant from the Abdus Salam International Centre for Theoretical Physics (ICTP), Trieste, Italy.  H.A.Z also acknowledges for the financial support of the National Scholarship Programme of the Slovak Republic (N{\v S}P). Jozef Stre{\v c}ka acknowledges the financial support of the Slovak Research and Development Agency under Grant No. APVV-20-0150 and scientific grant agency by the Ministry of Education, Science, Research and Sport of the Slovak Republic under Grant No. VEGA 1/0531/19.

\end{document}